\documentclass[superscriptaddress,amsmath,amssymb,prd,preprintnumbers,showpacs,twocolumn,nofootinbib]{revtex4-2}
\usepackage[colorlinks=true,pdfstartview=FitV,linkcolor=blue,citecolor=blue,urlcolor=blue,breaklinks=true]{hyperref}
\usepackage[T1]{fontenc} % if needed
\usepackage{amssymb,amsmath,bm,natbib}
\usepackage{color}
\usepackage{slashed}
\usepackage{graphics}
\usepackage{graphicx}
\usepackage[utf8]{inputenc}
\usepackage[caption=false]{subfig}
\usepackage{hyperref}
\usepackage{url}
\usepackage{dsfont}
\usepackage{float}
\usepackage{cancel}
\usepackage{units}
\usepackage{blindtext}
\usepackage[utf8]{inputenc}
\usepackage{upgreek}
\usepackage{booktabs}
\usepackage[dvipsnames,table,xcdraw]{xcolor}
\usepackage{enumerate}
\usepackage{mathtools}
\usepackage{soul,color}
\usepackage[normalem]{ulem}
\newcommand{\orcid}[1]{\href{https://orcid.org/#1}{\includegraphics[width=10pt]{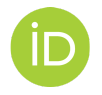}}}

\begin{document}
\title{G\"{o}del-symmetric backgrounds and explicit spacetime symmetry breaking}
\author{C\'esar Riquelme\orcid{0000-0003-0837-3891}}
\email{ceriquelme@udec.cl}
\affiliation{Departamento de F\'{i}sica, Universidad de Concepci\'{o}n, Concepci\'on, Casilla 160-C, Chile}
\affiliation{Centro de Ciencias Exactas, Facultad de Ciencias, Universidad del B\'{i}o-B\'{i}o, Chill\'{a}n, Casilla 447, Chile}
\author{Carlos M. Reyes\orcid{0000-0001-5140-6658}}
\email{creyes@ubiobio.cl}
\affiliation{Centro de Ciencias Exactas, Facultad de Ciencias, Universidad del B\'{i}o-B\'{i}o, Chill\'{a}n, Casilla 447, Chile}
\author{A. F. Santos\orcid{0000-0002-2505-5273}}
\email{alesandroferreira@fisica.ufmt.br}
\affiliation{Programa de Pós-Graduação em Física, Instituto de Física, Universidade Federal de Mato Grosso, Cuiabá, Brasil}
\begin{abstract}
Explicit breaking of diffeomorphism symmetry with nondynamical 
background fields in gravitational theories  
can lead to inconsistencies between the equations of motion and 
the underlying pseudo-Riemannian geometry. These theories 
produce a constraint equation that follows from the modified Einstein's 
equations and the
contracted Bianchi identity which has no simple solution and raises questions about 
 the consistency of backgrounds structures in extensions to general relativity.
  In contrast, spontaneous symmetry breaking
where fields 
 acquire a vacuum expectation value are shown to
 avoid these problems.
Nevertheless, there are some approaches that consistently implement 
 explicit diffeomorphism breaking by:
i) introducing geometrical restrictions which may 
dynamically restore diffeomorphism invariance,
ii) using the St\"{u}ckelberg formalism where auxiliary scalar fields can 
 carry degrees of freedom associated to spontaneously broken symmetries,
  and more recently by iii)  exploiting the isometries of a given gravitational configuration in the presence of a background;
  which ultimately requires the background fields to be Lie-dragged along the Killing vectors.
  In this work, we further develop and apply the latter approach focusing on a concrete example involving the
  G\"{o}del metric and the minimal gravitational Standard-Model Extension.
 We show that the resulting dynamics is fully consistent with the constraint equation and produce new 
 Noether's identities.
Furthermore we investigate causality violations by analyzing the critical 
radius, which is found to depend explicitly on the background fields. This dependence 
allows for the emergence of both causal and non-causal regions in spacetime.
\end{abstract}
\maketitle
%...................................
\section{Introduction}
\label{sec:introduction}
%...................................
Gravitational effective theories have become increasingly important due to their ability 
to describe phenomena beyond General Relativity (GR) and, in particular, beyond 
the standard $\Lambda$CDM cosmological model. Over recent decades, the latter 
has been confronted with a vast body of observational data testing its underlying 
assumptions about dark matter and dark energy. Yet the mechanism responsible
for the current acceleration of the Universe remains puzzling, especially in light 
of the extremely small value of the cosmological constant and recent data suggesting 
a dynamical cosmological parameter~\cite{DESI:2025zgx,DESI:2025fii}.
Despite extensive experimental searches, the fundamental nature of dark matter is still unknown. Additional hints of possible new 
physics are provided by observations suggesting the presence of birefringence in 
the CMB~\cite{Minami:2020odp,Lue:1998mq,Eskilt:2022cff}. Collectively, these 
open issues highlight the importance of exploring alternative theories of gravity.
  
 Several modified gravity theories have been put forward to deal with these challenges, including
 massive gravity~\cite{deRham:2010kj}, Einstein–Aether~\cite{Jacobson:2000xp,Jacobson:2004ts}, 
 Ho\v{r}ava–Lifshitz gravity~\cite{Horava:2008ih,Horava:2009uw}, and Chern–Simons gravity~\cite{Jackiw:2003pm}.
 The latter explicitly introduces the breaking of diffeomorphism invariance, which constitutes the main ingredient 
 of the modified gravitational framework considered in this work.
In general, two main approaches can 
be distinguished in the context of local Lorentz and diffeomorphism invariance
 breaking. The first introduces a fixed background structure which, although 
coordinate-dependent, remains non-dynamical in the sense that it does not 
fluctuate under perturbations 
of the metric or other dynamical variables. The second corresponds to spontaneous 
symmetry 
breaking, in which a dynamical field acquires a vacuum expectation value (VEV). For 
suitable choices of 
potential terms in the action, the resulting vacuum solutions can ``freeze''  into 
configurations that break diffeomorphism invariance.
   
The effective Standard-Model Extension (SME) provides a comprehensive, model-independent
 framework for investigating suppressed effects of local Lorentz and diffeomorphism 
 symmetry breaking~\cite{Colladay:1996iz,Colladay:1998fq}. The gravitational 
 sector~\cite{Kostelecky:2003fs} extends GR through spacetime-dependent coefficients 
 that couple to geometric operators. A particularly well-defined and widely studied subset
  is the minimal gravitational SME, which contains all power-counting renormalizable 
  operators, i.e., those of mass dimension four or less. More general extensions with 
  arbitrary operator dimensions have also been developed~\cite{Kostelecky:2017zob,Kostelecky:2020hbb}.
 
 A key insight into the nature of spacetime symmetry breaking within the SME, and in fact in any coordinate-independent 
 Lorentz-violating framework, lies in the distinction between observer and particle
  transformations. Observer transformations
are symmetries of the theory, whereas 
 particle transformations, being active transformations of the dynamical fields, can break them. 
 This distinction is fundamental in formulating a consistent framework for spacetime symmetry violation.
 It is also customary to distinguish between explicit breaking and spontaneous breaking, as mentioned above.
Explicit symmetry breaking is produced 
by a fixed nondynamical background, while spontaneous breaking arise due to a VEV
 acquired by the dynamical fields; giving rise to Nambu–Goldstone and massive 
modes~\cite{Bluhm:2004ep,Bluhm:2007bd}.
In the spontaneous case, diffeomorphism symmetry is preserved, although the solutions can depend 
on background structures originating from a potential term. Modified gravity theories such as the vector bumblebee 
gravity~\cite{Kostelecky:2003fs} or frameworks involving the Kalb–Ramond tensor field~\cite{Altschul:2009ae} 
provide examples where the Noether symmetries associated to diffeomorphism remain intact.
Within the context of alternative gravity theories 
and the gravitational sector of the SME, several aspects have 
been investigated, including Finsler geometries~\cite{Kostelecky:2011qz,AlanKostelecky:2012yjr,Schreck:2015seb,Edwards:2018lsn},
formal extensions~\cite{Davis:2025die},
the post-Newtonian limit~\cite{Bailey:2006fd}, Hamiltonian 
analysis~\cite{ONeal-Ault:2020ebv,Reyes:2021cpx,Reyes:2022mvm}, and boundary terms~\cite{Reyes:2023sgk}. Moreover, 
stringent experimental bounds on the effective coefficients have been obtained~\cite{Kostelecky:2008ts}. 

 It was recognized early on that explicit violation of diffeomorphism invariance can lead to 
 inconsistencies between the gravitational dynamics and the underlying Riemannian 
 geometry~\cite{Kostelecky:2003fs,Bluhm:2014oua,Bluhm:2019ato,Bluhm:2016dzm}. In particular, reconciling
  the modified Einstein field equations with the contracted Bianchi identities yields a coupled system of 
  partial differential equations involving the SME background coefficients, which is in general highly 
  nontrivial to solve. This fundamental difficulty is often referred to as the ``no-go” constraint, 
  highlighting the challenges in constructing consistent gravitational models with explicit symmetry breaking.
Several strategies have been developed to address these issues. One approach 
restricts the space of metric solutions of the modified Einstein equations in such 
a way that the Noether identities remain satisfied; in some cases, this results in a 
dynamical restoration of diffeomorphism invariance. A notable example is the gravitational 
Chern–Simons theory~\cite{Jackiw:2003pm}, where consistency is maintained by
 imposing the condition ${}^{*}R^{\sigma\phantom{\tau}\mu\nu}_{\phantom{\sigma}\tau}
 R^{\tau}_{\phantom{\tau}\sigma\mu\nu}=:{}^{*}RR=0$, where $R^{\tau}_{\phantom{\tau}\sigma\mu\nu}$ is
  the Riemann curvature tensor and ${}^{*}R^{\tau\phantom{\sigma}\mu\nu}_{\phantom{\tau}\sigma}:
  =(1/2)\varepsilon^{\mu\nu\alpha\beta}R^{\tau}_{\phantom{\tau}\sigma\alpha\beta}$ is its dual.
  More recently, within generalizations of the minimal SME, suitable combinations 
  of symmetry-breaking terms have been shown to preserve the Noether
   identities~\cite{Bailey:2024zgr}, albeit at the cost of introducing discontinuities in the linearized regime.
  Another widely employed approach
   is the St\"{u}ckelberg formalism, in which scalar fields $\phi^A$, with internal indices $A$, 
   are introduced to reproduce the degrees of freedom associated to broken symmetries~\cite{Bluhm:2023kph}.

A recent approach to addressing the no-go constraint exploits the isometries of a given gravitational 
configuration~\cite{Reyes:2024ywe,Reyes:2024hqi,Reyes:2022dil}. Within this framework, the energy–momentum tensor associated with the 
background fields is required to remain invariant under the transformations generated 
by the system of Killing vectors. This condition, in particular applied in cosmology, 
 introduces additional symmetry 
directions along which the Bianchi identities can be satisfied, thereby helping to
 circumvent the restrictions imposed by observer diffeomorphism invariance. Although 
 this strategy constrains the background and reduces the number of independent 
 components, it improves the prospects for achieving compatibility between the Bianchi
  identities and the modified equations of motion. A natural consequence of this requirement
   is that the background fields themselves must be invariant under the isometries of the 
   spacetime. Hence, any consistent formulation of explicit diffeomorphism violation must 
   ensure that both the Bianchi identities and the equations of motion are satisfied simultaneously. 
   As a concrete example, we extend the formalism to consider the G\"{o}del metric and explore the consistency of solutions within 
   the gravitational sector of the SME.

The G\"{o}del metric is an exact solution to 
 Einstein’s field equations in GR~\cite{Godel, Obukhov, Reboucas}. It describes a stationary, rotating 
 universe and is particularly notable for admitting closed timelike curves (CTCs), which theoretically 
 allow a traveler to return to their own past, raising profound questions about the nature of 
 causality. Importantly, this causality violation arises from the global structure of 
 spacetime, while locally, GR remains consistent with the principles of 
 special relativity, preserving causality in small regions. However, the presence of CTCs is 
 not unique to the G\"{o}del solution. Such features appear in a range of solutions to 
 Einstein's equations, including the Kerr black hole~\cite{Kerr:1963ud}, the van 
 Stockum spacetime~\cite{Van}, and models involving cosmic strings~\cite{Gott:1990zr}, among others. 
 These examples suggest that the emergence of CTCs is a broader feature of GR under 
 certain conditions rather than an anomaly exclusive to G\"{o}del’s universe.

In recent years, various extensions of GR have been explored to study how these exotic structures 
behave in modified gravitational frameworks. Notably, the G\"{o}del solution has been investigated
 in the context of Bumblebee gravity, a model that incorporates spontaneous Lorentz symmetry 
 breaking and represents the gravitational sector of the SME~\cite{santos2015godel, jesus2020godel}. 
 In the present work, the G\"{o}del metric is analyzed within the minimal gravitational 
 sector of the SME, where explicit violations of diffeomorphism and Lorentz invariance 
 are introduced. The associated Killing vectors are computed, and the corresponding 
 G\"{o}del invariants are determined. These modifications lead to changes in the field equations, 
 which are then solved to examine the consistency of the G\"{o}del solution
 in this theoretical framework. Special attention is given to how the background fields
  affect both the existence of the solution and the conditions under which causality is violated.

The present paper is organized as follows. 
 In Sec.~\ref{sec:explicit_diff}, we review the potential problems and peculiar features 
 associated with explicit spacetime symmetry breaking. To this end, we discuss Noether’s
  theorem in the presence of background fields. We then present a toy model in Minkowski 
  spacetime that illustrates some of the issues that arise when gravity is subsequently included.
In Sec.~\ref{sec:sme}, we present an overview of the minimal SME with explicit 
diffeomorphism violation. Sec.~\ref{Godel} introduces the G\"{o}del metric, discusses
 its symmetries, and presents 
 the G\"{o}del form-invariant tensors, including rank-0 scalars, symmetric rank-2 tensors, 
 and Riemann-like rank-4 tensors. Sec.~\ref{sec:implementation} is devoted to analyzing 
 the conditions under which the G\"{o}del metric arises as a solution within the gravitational 
 SME. In Sec.~\ref{causality}, we investigate causality violation in each SME sector by 
 evaluating the critical radius. Finally, Sec.~\ref{sec:conclusions} contains our concluding remarks.
%............................................................................................
 \section{Backgrounds and 
Noether’s theorem}\label{sec:explicit_diff}
%............................................................................................
Background fields have a profound impact on the formulation of Noether’s theorem. When 
such fixed structures are included in the action, the breaking of global symmetries leads to 
background-dependent relations that replace the usual conserved-current equations, while 
the breaking of local symmetries results in modified Noether identities. In what follows, we 
first review Noether’s theorem and then analyze a CPT-odd, Lorentz-violating extension of 
electrodynamics with a spacetime-dependent background field. By using this electromagnetic 
toy model we draw a parallel with explicit diffeomorphism breaking in gravity. Finally, we 
show how imposing spacetime isometries can be used to address the associated consistency issues.
%............................................................................................
 \subsection{Noether’s theorem}\label{subsec:explicit_diff}
 %...........................................................................................
 %............................................................................................
 \subsubsection{Global symmetries}\label{subsubsubsec2}
 %...........................................................................................
 Consider the action for a set of scalar fields $\phi_I(x)$,
\begin{align}
S[\phi_I]=&\int  \mathrm{d}^4x\, \mathcal L\left(\phi_I,\partial_{\mu }\phi_I\right)  \,,
\end{align} 
where the index $I$ labels components in an internal symmetry group $G$.

The Euler-Lagrange 
equations of motion are
\begin{align}\label{EL_eq}
\frac{\partial \mathcal L}{\partial  \phi_I}  - 
 \partial_{\mu}\left(\frac{\partial \mathcal L}{\partial\left( \partial_{\mu} \phi_I \right)} \right)  =0  \,.
\end{align}

We now introduce a symmetry transformation of the fields, $\delta_s\phi_I$, 
defined such that the action is invariant up to a total derivative or
a boundary term.
By introducing the notation for the variation of the action 
\begin{align}\label{Def_var}
\delta S[\phi_I, \delta _s \phi_I]:=&   S[\phi_I+ \delta _s \phi_I]- S[\phi_I]   \,,
\end{align}
we have under a symmetry transformation
\begin{align}
\delta S[\phi_I,\delta_s\phi_I]
= \int \mathrm{d}^4x\, \partial_{\mu}\Lambda^{\mu} \,.
\end{align}

Consider the infinitesimal 
 coordinate transformation 
  \begin{align}\label{diff_tranf}
x^{\mu}\to x^{\mu}+\varepsilon^{\mu}  \,,
\end{align}
with $\varepsilon^{\mu}$ an infinitesimal four-vector. The field transforms as
  \begin{align}
\delta \phi_I =-\varepsilon^{\lambda} \partial_{\lambda} \phi_I \,,
\end{align}
 and since $\mathcal L$ is a scalar under~\eqref{diff_tranf}, one finds
\begin{align}\label{varS1}
\delta S[\phi_I,\delta_s\phi_I]
=-\!\int \mathrm{d}^4x\,\varepsilon^{\lambda}\partial_{\lambda}\mathcal L
= -\!\int \mathrm{d}^4x\,\partial_{\lambda}(\varepsilon^{\lambda}\mathcal L) \, .
\end{align}
Now we consider an arbitrary variation of the field which 
we denote as usual as $\delta\phi_I$. The action varies as
 \begin{align}\label{var_general}
\delta S[  \phi_I, \delta \phi_I]=&\int \mathrm{d}^4x  \Bigg \{  
 \left(  \frac{\partial \mathcal L}{\partial  \phi_I}  -  
  \partial_{\mu}\left(\frac{\partial \mathcal L}{\partial\left( \partial_{\mu} \phi_I \right)} \right)  
    \right) \delta \phi_I \notag  \\ & + \partial_{\mu}\left( \frac{\delta \mathcal L}{\partial (\partial_{\mu} \phi_I)} \delta \phi_I \right) \Bigg \}  \,.
\end{align}
 Evaluating \eqref{var_general} on solutions of the equations of 
 motion, which we denote by $\bar\phi_I$, the bulk term vanishes and one obtains
\begin{align}\label{varS2}
\delta S[\bar  \phi_I, \delta \phi_I]=&\int \mathrm{d}^4x\, \partial_{\mu}
\left( \frac{\delta \mathcal L}{\partial (\partial_{\mu} \phi_I)} \delta \phi_I \right)\,,
\end{align}
where the variation is arbitrary and
called on-shell variation, see~\cite{Banados:2016zim,Fleming:1987up,Jackiw:1986dr}.

Finally, considering $\phi_I=\bar  \phi_I$ in~\eqref{varS1}
and 
$\delta  \phi_I=\delta _s \phi_I$ in~\eqref{varS2} 
 and 
subtracting both terms,
we arrive at 
\begin{align}
0=&\int \mathrm{d}^4x\, \partial_{\lambda} J^{\lambda}  \,,
\end{align}
with the Noether conserved current
\begin{align}
J^{\lambda}=& -\varepsilon^{\lambda}\mathcal L  +  \frac{\delta \mathcal L}{\partial (\partial_{\lambda} \phi_I)}
 \varepsilon^{\alpha} \partial_{\alpha} \phi_I  \,.
\end{align}
 %............................................................................................
 \subsubsection{Local symmetries}\label{subsubsubsec:3}
 %...........................................................................................
As is well known, for local symmetries such as diffeomorphisms in GR, Noether’s procedure yields
identities rather than conserved currents. To illustrate this, we focus on the Einstein–Hilbert (EH) action,
\begin{align}
S[g_{\mu \nu}]=&\frac{1}{2\kappa} \int \mathrm{d}^4x\, \sqrt{-g} R \,.
\end{align}
For an infinitesimal diffeomorphism generated by $\xi^\mu(x)$,
\begin{align}\label{GC}
    x'^{\mu}&=x^{\mu}+\xi^{\mu}(x) \,,
\end{align}
the variation of the scalar density $\sqrt{-g}\,R$ reads
\begin{align}
\delta (\sqrt{-g} R) &= \delta  (\sqrt{-g} ) R+\sqrt{-g} \, \delta R   \,.
\end{align}
Using standard identities, we can write
\begin{align}
\delta(\sqrt{-g} R)= \frac 12 \sqrt{-g}  g^{\alpha \beta} \delta g_{\alpha \beta}  R+\sqrt{-g}  \, \delta R  \,.
\end{align}
If we consider the variation induced by the generalized translation~\eqref{GC}, we have
\begin{align}
\delta_{\xi} g_{\alpha \beta}  =-(\nabla_{\alpha} \xi_{\beta}+ \nabla_{\beta} \xi_{\alpha} ) \,,
\end{align}
and also
\begin{align}
\delta_{\xi} R =- \xi^{\alpha} \nabla_{\alpha}R  \,,
\end{align}
therefore we obtain
\begin{align}
\delta_{\xi}(\sqrt{-g} R)= -\sqrt{-g}  \nabla_{\mu}(  \xi^{\mu}  R) \,.
\end{align}

The variation of the EH action with $\xi$ nonzero on the boundary is given by
\begin{align}\label{var1}
    \delta S[g_{\mu \nu}, \delta_{\xi} g_{\mu \nu}]=-\frac{1}{2\kappa} \int\mathrm{d}^4x \sqrt{-g} \nabla_{\mu}( \xi^{\mu} R)\,,
\end{align}
where we have used the definition~\eqref{Def_var}. 

Now, let us take an arbitrary variation of the action with respect to the 
 fields, i.e., the metric field
\begin{align}\label{var2}
    \delta S[g_{\mu \nu}, \delta g_{\mu \nu}]&=\frac{1}{2\kappa}  \int\mathrm{d}^4x  \sqrt{-g} \bigg[ \left(R_{\mu \nu}-\frac 12 g_{\mu \nu} R\right)  \delta g^{\mu \nu}
   \notag  \\ &\hspace{2em}+   P^{\mu \nu \alpha \beta} \nabla_{\mu} \nabla_{\nu} \delta g_{\alpha \beta} \bigg]   \,,
\end{align}
where we have used
\begin{align}
   g^{\mu \nu} \delta R_{\mu \nu}= P^{\mu \nu \alpha \beta} \nabla_{\mu} \nabla_{\nu}
   \delta g_{\alpha \beta} \,,
\end{align}
with the definition
\begin{align}\label{def_P}
    P^{\mu\nu\alpha\beta}=g^{\mu \alpha} g^{\nu \beta}  - g^{\mu \nu} g^{\alpha \beta}   \,.
\end{align}
Setting both variations to be the same, i.e.,
$\delta g_{\mu \nu}= \delta_{\xi} g_{\mu \nu}$
and subtracting Eqs.~\eqref{var1} and \eqref{var2}, we obtain 
\begin{align}\label{int_var}
 0&=\int\mathrm{d}^4x  \sqrt{-g} \bigg[-\nabla_{\mu}( 
 \xi^{\mu}  R) -G_{\mu \nu} \delta_{\xi} g^{\mu \nu}
 \notag  \\&  -   P^{\mu \nu \alpha \beta} \nabla_{\mu}
 \nabla_{\nu} \delta_{\xi} g_{\alpha \beta} \bigg]   \,,
\end{align}
where we have introduced the Einstein tensor
$  G_{\mu \nu}= R_{\mu \nu}-\frac 12 g_{\mu \nu} R $.
Now, with 
\begin{align}
  \delta_{\xi} g^{\mu \nu} =  \nabla^{\mu}\xi^{\nu}+\nabla^{\nu}\xi^{\mu}  \,,
\end{align}
and manipulating the middle term,
\begin{align}
 G_{\mu \nu}  \delta g^{\mu \nu} =  G_{\mu \nu}  ( \nabla^{\mu}\xi^{\nu}+\nabla^{\nu}\xi^{\mu})
 = 2 \nabla_{\mu}(G^{\mu}_{\phantom{\mu} \nu}  \xi^{\nu})    \,,
\end{align}
where we have used the contracted Bianchi identity \(\nabla_{\mu}G^{\mu}{}_{\nu}=0\), arrive at 
\begin{align}
0&=\int\mathrm{d}^4x  \sqrt{-g} \bigg[-\nabla_{\mu}(\xi^{\mu} R) -2 \nabla_{\mu}(G^{\mu}_{\phantom{\mu} \nu}  \xi^{\nu})
  \notag \\& +   P^{\mu \nu \alpha \beta} \nabla_{\mu} \nabla_{\nu} (\nabla_{\alpha}\xi_{\beta}+\nabla_{\beta}\xi_{\alpha} ) \bigg]   \,.
\end{align}
Cleaning a little bit, and 
rearranging the derivative term into a total divergence
\begin{align}
0&=\int\mathrm{d}^4x  \sqrt{-g}\nabla_{\mu}  \bigg[ -2 (R^{\mu}_{\phantom{\mu} \nu}  \xi^{\nu})
  \notag \\& +   P^{\mu \nu \alpha \beta}  \nabla_{\nu} (\nabla_{\alpha}\xi_{\beta}+\nabla_{\beta}\xi_{\alpha} ) \bigg]   \,,
\end{align}
takes the standard Noether form
\begin{align}
 0&=\int\mathrm{d}^4x  \sqrt{-g}\; \nabla_{\mu}  J^{\mu}_K   \,,
\end{align}
with
\begin{align}
 J_K^{\mu}:=  R^{\mu}_{\phantom{\mu} \nu}  \xi^{\nu}
- \frac 12   P^{\mu \nu \alpha \beta}  \nabla_{\nu} (\nabla_{\alpha}\xi_{\beta}+\nabla_{\beta}\xi_{\alpha} )    \,.
\end{align}
or
\begin{align}\label{Kom_current}
 J^{\mu}_K(\xi)=  \nabla_{\nu} (\nabla^{[\mu}\xi^{\nu] }  )    \,.
\end{align}
The above piece is the Komar current,
which satisfies an identically current conservation relation 
\begin{align}\label{KCurrent_cons}
\nabla_{\mu} J_K^{\mu}(\xi)\equiv  0  \,.
\end{align}
%................................................................................................................
 \subsection{Modified Maxwell theory with a spacetime-dependent term $\bar {k}^{\mu}(x)$}\label{subsec1:explicit_diff}
%................................................................................................................
In this subsection, we present a parallel with explicit 
symmetry breaking in gravity. To this end, we consider a CPT-odd 
term in the SME photon sector with a spacetime-dependent background.

Consider the modified Maxwell action
\begin{align}
S[A_{\mu}]=\int \mathrm{d}^4x\,   {\mathcal L}_{\text{MCS},k}  \,,
\end{align}
with Lagrangian density
\begin{align}
 {\mathcal L}_{\text{MCS},k} = -\frac 14 F_{\mu \nu} F^{\mu \nu}
+ \frac 12  \bar {k}^{\mu}(x) A^{\alpha} \epsilon_{\mu \alpha \beta \gamma} F^{\beta \gamma}  \,.
\end{align} 
The first term is the usual Maxwell term and the second 
a like Chern-Simons term with a spacetime-dependent background 
 \({\bar k}^{\mu}(x)\); for a constant coefficient see Ref.~\cite{Colladay:1998fq}. Also
\(F_{\mu\nu}=\partial_{\mu}A_{\nu}-\partial_{\nu}A_{\mu}\) is the electromagnetic field strength,
and \(\epsilon_{\mu\alpha\beta\gamma}\) denotes the Levi--Civita tensor.

Let us consider the variations
\begin{align}
\frac{  \partial    {\mathcal L}_{\text{MCS},k}    }{\partial  A_\alpha}
= \frac 1 2  \bar{k}^\mu \eta^{\lambda \alpha} \epsilon_{\mu \lambda \beta \gamma}F^{\beta \gamma }\,,
\end{align}
and
\begin{align}
\frac{\partial {\mathcal L}_{\text{MCS},k}     }{\partial (\partial_\mu A_\alpha)}
= -F^{\mu\alpha} + \bar{k}^\nu A^\lambda   
\eta^{\mu \delta}\eta^{\alpha \epsilon} \epsilon_{\nu\lambda\delta \epsilon}\,.
\end{align}
The Euler-Lagrange equations of motion are
\begin{align}\label{E-L-eq-k}
    \partial_\mu F^{\mu\alpha} &+\eta^{\alpha\nu}\bar{k}^{\mu}\epsilon_{\mu\nu\beta\gamma} 
    F^{\beta\gamma} \notag \\ &\hspace{3em} 
    -\eta^{\gamma\alpha}   
(\partial^\mu \bar{k}^{\nu})A^\lambda \epsilon_{\nu\lambda \mu\gamma}=0\,.
\end{align}
Taking the derivative $\partial_{\alpha}$ of Eq.~\eqref{E-L-eq-k}, we arrive at a
restriction involving both the background and the electromagnetic tensor. We can call
the subsequent relation the  
no-go condition in this model
\begin{align}\label{type-nogo}
 (\partial^\mu \bar{k}^{\nu})\, \epsilon_{\nu\mu\alpha\lambda} F^{\alpha \lambda}  =0 \,.
\end{align}
Considering a global transformation~\eqref{diff_tranf} the gauge field transforms as
\begin{equation}\label{Global_S3}
\delta_s A_{\mu}=-\varepsilon^{\lambda}\partial _{\lambda} A_{\mu}  \,,
\end{equation}
and since the Lagrangian is a scalar under such transformation, we have
\begin{equation}\label{Global_S2}
\delta S[A_{\mu}, \delta _s A_{\mu}]=-\int \mathrm{d}^4x\, 
\varepsilon^{\lambda} \partial_{\lambda} {\mathcal L}_{\text{MCS},k} \,.
\end{equation}
Now we consider the variation of the action with respect to the fields,
\begin{align}
& \delta S [A_\mu,\delta A_{\mu}]=\; \int d^4x \,\Bigg[
\Bigg( \frac{\partial {\mathcal L}_{\text{MCS},k}   }{\partial A_\nu}
- \partial_\rho \frac{\partial {\mathcal L}_{\text{MCS},k}    }{\partial (\partial_\rho A_\nu)} \Bigg) \delta A_\nu
 \notag \\&\hspace{1cm}+ \frac{\partial {\mathcal L}_{\text{MCS},k}   }{\partial \bar{k}^\mu}\,\delta \bar{k}^\mu
+ \partial_\rho \Bigg(
\frac{\partial {\mathcal L}_{\text{MCS},k}   }{\partial (\partial_\rho A_\nu)} \,\delta A_\nu
\Bigg) \Bigg]\,.
\end{align}
Evaluated on shell with $\bar{A}_\mu$, the Euler-Lagrange term vanishes and taking
 $\delta A_{\nu}=-\varepsilon^{\lambda} \partial_{\lambda}A_{\nu}$ 
and  $\delta {\bar k}^{\nu}=-\varepsilon^{\lambda} \partial_{\lambda}  {\bar k}^{\nu}$,
we obtain
\begin{align}\label{VarS}
\delta S [\bar{A}_\mu,\delta_s A_{\mu}]& =-\int d^4x \,   \varepsilon^{\lambda}   \Bigg[    
  \frac{\partial {\mathcal L}_{\text{MCS},k}  }{\partial \bar{k}^\mu}  \partial_{\lambda} {\bar k}^{\nu}  \notag \\&\hspace{1cm}+ \partial_\rho \Bigg(
\frac{\partial {\mathcal L}_{\text{MCS},k}   }{\partial (\partial_\rho A_\nu)} \partial_{\lambda}A_{\nu}
\Bigg) \Bigg] \,.
\end{align}
Notice that the variation~\eqref{VarS} is not 
a total derivative and hence we do not have a conserved current according to 
the Noether theorem. However we may still extract information of the above equation.

As before we consider an on-shell field in~\eqref{Global_S2} 
and subtract with~\eqref{VarS}, arriving at the relation 
\begin{align}\label{Almost_Noether}
-\partial_{\lambda} {\mathcal L}_{\text{MCS},k}   &+\partial_{\mu}\Bigg( \frac{\partial {\mathcal L}_{\text{MCS},k} }{\partial (\partial_\mu A_\alpha)} 
\,\partial_{\lambda} A_\alpha\Bigg)\notag \\ &\hspace{2.5em}
+ 
\frac{\partial {\mathcal L}_{\text{MCS},k} }{\partial \bar{k}^\mu}  \partial_{\lambda} \bar{k}^\mu=0\,.
\end{align}
Using 
\begin{align}
  \frac{\partial {\mathcal L}_{\text{CS},k} }{\partial \bar{k}^\mu}=
  \frac 12 \epsilon_{\mu\alpha\beta\gamma} A^{\alpha}F^{\beta\gamma} \,,
\end{align}
and the definition of the would-be conserved 
energy-momentum tensor $T^{\nu}_{\phantom{\nu}\lambda}$,
replaced in Eq.~\eqref{Almost_Noether}, we obtain
\begin{align}
 \partial_{\nu}T^{\nu}_{\phantom{\nu}\lambda} =-  
 \frac 12 \epsilon_{\mu\alpha\beta\gamma} A^{\alpha}F^{\beta\gamma} \partial_{\lambda} \bar{k}^\mu\,.
\end{align}
This relation shows that energy-momentum conservation is lost unless
the right-hand side vanishes. 
In particular, for constant $\bar {k}^{\mu}$ the standard
conservation law is recovered~\cite{Colladay:1998fq}. Other nontrivial possibilities also exist. For instance,
if the background is a pure gradient,
\begin{align}
{\bar k}^{\mu}(x)=\partial^{\mu} \phi(x)\,,
\end{align}
the no-go condition~\eqref{type-nogo} is satisfied 
and in this case the action modulo total derivatives
can be rewritten as 
\begin{align}
S[A_{\mu}]=\int d^4 x   \left( -\frac 14 F_{\mu \nu} F^{\mu \nu}
- 2 \phi(x) F^{\mu \nu} \tilde F_{\mu \nu}   \right)\,,
\end{align}
with $\tilde F_{\mu \alpha} =\frac 12 \epsilon_{\mu\alpha\beta\gamma} F^{\beta\gamma}  $.

This procedure introduces the Pontryagin density through an axion-like coupling. Such terms
have been extensively studied in axion physics, see the review~\cite{Irastorza:2021tdu} and in the context of topological
insulators~\cite{Hasan:2010xy}. Moreover, the appearance of the field $\phi(x)$ resembles a
St\"uckelberg-type construction, in which an additional degree of freedom is introduced
to restore gauge invariance.
%............................................................................................
 \subsection{Backgrounds in gravity}\label{subsec2:explicit_diff}
%............................................................................................
Finally, we turn to gravity in the presence of background fields.
In light of the previous construction, we aim to show how the 
no-go constraint arises~\cite{Bluhm:2014oua} and how invoking spacetime
isometries provide a consistent resolution~\cite{Reyes:2024ywe}.

Let us focus on the effective gravitational action
\begin{align}\label{eq:modified-action}
 S'_g[g_{\mu \nu}]&=\int_{\mathcal{M}} \mathrm{d}^4x\,\frac{\sqrt{-g}}{2\kappa}\,\Big(R
+\mathcal{L}'(g_{\mu\nu},\bar{k}^{\alpha\beta\dots\omega})\Big)\,,
\end{align}
where we have introduced the coefficients $\bar{k}^{\alpha\beta\dots\omega}$ of a generic nondynamical
background field. We are considering boundary contributions in $S_g'$ 
in order to define a consistent variational formalism~\cite{Reyes:2023sgk}.

Under a variation of the metric we arrive at 
the modified Einstein's equation
\begin{align}
\label{eq:modified-action}
-G_{\mu \nu}+  T'_{\mu \nu}&=0\,,
\end{align}
where 
\begin{equation}
T'^{\mu\nu}:= \frac{1}{\sqrt{-g}} \frac{\delta(\sqrt{-g}\mathcal{L}')}{\delta g_{\mu\nu}}\,.
\end{equation}
We follow the notation and method of the previous subsections
and consider a local coordinate transformation~\eqref{GC}. Hence, we have the symmetry transformation
\begin{align}\label{eq:modified-action}
\delta S_g'[g_{\mu \nu}, \delta_{\xi} g_{\mu \nu}]&=-\int\mathrm{d}^4x\,\,
\frac{\sqrt{-g}}{2\kappa} \nabla_{\mu}  \bigg[ \xi^{\mu}
\Big(R \notag \\ &\hspace{3em}
+\mathcal{L}'(g_{\mu\nu},\bar{k}^{\alpha\beta\dots\omega})\Big)  \bigg]\,.
\end{align}
Now, a variation of the fields including the background produces
\begin{align}\label{Obs_trans}
\delta S_g'[g_{\mu \nu}, \delta g_{\mu \nu}]&=\int\mathrm{d}^4x\,\frac{\sqrt{-g}}{2\kappa}  \Big[   \left(-G^{\mu\nu}+ 
T'^{\mu\nu}\right)\delta g_{\mu\nu}  \\ &  + P^{\mu \nu \alpha \beta} \nabla_{\mu} \nabla_{\nu} \delta g_{\alpha \beta}   +
J_{\alpha\beta\dots\omega}\delta \bar{k}^{\alpha\beta\dots\omega}     \Big]  \notag \,,
\end{align}
where we have cancel boundary terms arising from 
 $\mathcal L'$, and we have defined
\begin{equation}
J_{\alpha\beta\dots\omega}:=\frac{\delta\mathcal{L}'}{\delta\bar{k}^{\alpha\beta\dots\omega}}\,.
\end{equation}

We consider the induced variations $\delta g_{\mu \nu} =\delta_{\xi} g_{\mu \nu}$ and

$ \delta   \bar{k}^{\alpha\beta\dots\omega}= -\mathcal L_{\xi}   \bar{k}^{\alpha\beta\dots\omega}$ replaced in~\eqref{Obs_trans},
where $\mathcal L_{\xi}$ is the Lie derivative. 

Now, subtracting~\eqref{eq:modified-action} and~\eqref{Obs_trans}, one has
\begin{align}
0&=\int\mathrm{d}^4x\,\frac{\sqrt{-g}}{2\kappa}\,
\bigg\{- \nabla_{\mu}  \Big[ \xi^{\mu}
\Big(R
+\mathcal{L}'(g_{\mu\nu},\bar{k}^{\alpha\beta\dots\omega})\Big)  \Big] \notag  \\ &\hspace{2em} -  \Big[      
\left(-G^{\mu\nu}+ 
T'^{\mu\nu}\right) \delta_{\xi} g_{\alpha \beta}  +P^{\mu \nu \alpha \beta} \nabla_{\mu} \nabla_{\nu} \delta_{\xi} g_{\alpha \beta}\notag  \\ &\hspace{2em} -
J_{\alpha\beta\dots\omega} \mathcal L_{\xi}   \bar{k}^{\alpha\beta\dots\omega}    \Big]\bigg\} \,.
\end{align}
After 
some calculation one can show that the expression above can be written as
\begin{align}\label{Eq_balance}
   0&=\int\mathrm{d}^4x \sqrt{-g}  \Big[
   J_{\alpha\beta\dots\omega}   \mathcal L_{\xi}   \bar{k}^{\alpha\beta\dots\omega}
   + 2 \xi_{\nu} \nabla_{\mu}  T'^{\mu\nu}\notag \\ &\hspace{3em}
   +  \nabla_{\mu}W^{\mu}(\xi, \bar{k}^{\alpha\beta\dots\omega}   )  \Big]   \,,
\end{align}
producing the new Noether identity
\begin{align}
  &  J_{\alpha\beta\dots\omega}   \mathcal L_{\xi}   \bar{k}^{\alpha\beta\dots\omega}
   + 2 \xi_{\nu} \nabla_{\mu}  T'^{\mu\nu}\notag \\ &\hspace{4em}
   +  \nabla_{\mu}W^{\mu}(\xi, \bar{k}^{\alpha\beta\dots\omega}   ) =0    \,,
\end{align}
Note that in the absence of explicit background one obtains the conservation of Komar current
Eq.~\eqref{KCurrent_cons}.
We expect 
\begin{align}\label{new_noether}
   \nabla_{\mu}W^{\mu}(\xi, \bar{k}^{\alpha\beta\dots\omega}   ) =0  \,,
\end{align}
where the quantity $W^{\mu}$ eventually depends on the model. We 
provide an example in the Appendix~\ref{Appendix} for the scalar sector $u$
of the SME.

Using~\eqref{new_noether}, we can write 
\begin{align}\label{Eq_balance2}
   0&=\int\mathrm{d}^4x \sqrt{-g}  \Big[
   J_{\alpha\beta\dots\omega}   \mathcal L_{\xi}   \bar{k}^{\alpha\beta\dots\omega}
   + 2 \xi_{\nu} \nabla_{\mu}  T'^{\mu\nu}
    \Big]   \,.
\end{align}
We see that in the directions where the Lie derivative of the background field vanishes,
we have
\begin{equation}\label{eq:consistency-requirement}
\nabla_{\mu}T'^{\mu \nu}=0\,,
\end{equation}
and we recover diffeomorphism invariance in these directions. 
However, along directions for which the Lie derivative of the background is non-vanishing, the 
symmetry is broken. In Noether language, this manifests itself in the fact that the new Noether identity does not produce a conserved 
quantity.

If we would take the four-vector $\xi$
to vanish on the boundary in Eq.~\eqref{Eq_balance2}, and perform repeated
integrations by parts, 
we would arrived at an identity for the background field coefficients
\begin{align}\label{eq:principal-equation-differential-form}
&2\nabla_{\mu}T'^{\mu}_{\phantom{\mu}\,\,\nu}=J_{\alpha\beta\dots\omega}
\nabla_{\nu}\bar{k}^{\alpha\beta\dots\omega}+
\nabla_{\lambda}(J_{\nu\beta\dots\omega}\bar{k}^{\lambda\beta\dots\omega})  \\
&\hspace{2em}   +\nabla_{\lambda}(J_{\alpha\nu\dots\omega}\bar{k}^{\alpha\lambda\dots\omega})
+\dots+\nabla_{\lambda}(J_{\alpha\beta\dots\nu}\bar{k}^{\alpha\beta\dots\lambda}) \notag \,,
\end{align}
which was derived using a different method in ~\cite{Reyes:2024ywe}.

In the next sections, we show that for a gravitational system that exhibits
Killing vectors fields it is possible to fulfill Eq.~\eqref{eq:consistency-requirement}.
In particular, we give examples using the G\"{o}del isometries.
%..............................................................................
\section{The gravitational SME} \label{sec:sme}
%....................................................................
The action describing the gravitational 
sector of the minimal SME with explicit diffeomorphism violation is given by
\begin{align}\label{1}
\tilde S&= \frac{1}{2\kappa}\int_{\mathcal{M}} \mathrm{d}^4x\,
\sqrt{-g}    \left(  R+2\Lambda+\mathcal L_{\text{SME}}   \right)  +S_{\partial \mathcal M}\notag  \\
&\hspace{2em}+S_m\,,
\end{align}
where the diffeomorphism
breaking contribution $\mathcal{L}_{\text{SME}}$ takes the form
\begin{align}
\mathcal L_{\text{SME}}&=-u {R}+s^{\mu \nu}{R}_{\mu \nu}+t^{\mu\nu\rho\sigma}{R}_{\mu\nu\rho\sigma}\,.
\end{align}
Here, $u$, $s^{\mu\nu}$, and $t^{\mu\nu\rho\sigma}$ denote non-dynamical 
background fields that explicit break diffeomorphism and 
 local Lorentz invariance. 
We have $\kappa = 8\pi G$, where $G$ is Newton’s gravitational constant, $g$
 is the determinant of the metric $g_{\mu\nu}$, $R$ is the Ricci scalar, $R_{\mu\nu}$ 
 is the Ricci tensor, $R_{\mu\nu\rho\sigma}$ is the Riemann tensor and $\Lambda$ 
 is the cosmological constant. 
 
 The term $S_{\partial \mathcal M}$ accounts for boundary contributions 
 required for a well-defined variational 
 principle~\cite{Reyes:2022mvm,Reyes:2022dil}, they can be written as 
  \begin{subequations}
 \begin{align}
S_{   \partial \mathcal{M}  }    ^{({u})}   &=-\oint_{\partial\mathcal{M}} \mathrm{d}^3y\,
  \varepsilon \frac{\sqrt{q}}{\kappa}\, uK   \,,
\\[1ex]
S_{   \partial \mathcal{M}  }    ^{({s})}   &=  \oint_{\partial\mathcal{M}} \mathrm{d}^3y\, 
 \varepsilon \frac{\sqrt{q}}{2\kappa}\,\left( s^{ab}K_{ab}-s^{\mathbf{nn}} K\right)  \,,
\\[1ex]
S_{   \partial \mathcal{M}  }    ^{({t})}   &= \oint_{\partial\mathcal{M}}  
 \mathrm{d}^3y \, \varepsilon \frac{\sqrt{q}}{\kappa} \,   2 t^{\mathbf{n}a\mathbf{n}b} K_{ab}   \,.
\end{align}
\end{subequations}
 Here, $q_{ab}$ is the induced metric on the boundary
 hypersurface $\partial \mathcal{M} $, $q$ its determinant, and
$\varepsilon=n_{\mu} n^{\mu}$ where $n_{\mu}$ is the boundary normal
evaluated at each point of $\partial\mathcal{M}$.
We have introduced
 the extrinsic curvature $K^{ab}$, its trace $K$, and 
the notation $X^{\bf {n}}:=n_{\mu }X^{\mu}$.
Furthermore, $S_m$ represents the action associated with the matter content.

By varying the action in Eq. (\ref{1}), the modified Einstein 
field equations for the gravitational sector of the SME are obtained
\begin{align}  \label{EM}
    G_{\mu\nu}+\Lambda g_{\mu\nu}&= (T^{Rstu})_{\mu\nu}+\kappa(T_m)_{\mu\nu} \,,
\end{align}
where $G_{\mu\nu}$ is the Einstein tensor, $(T_m)_{\mu\nu}$ denotes the 
energy-momentum tensor of the matter content, and the contribution from the 
Lorentz-violating background fields is represented by the rank-2 tensor
\begin{align}
   (T^{Rstu})^{\mu\nu}&=-\nabla^\mu \nabla^\nu u+g^{\mu\nu}\nabla^2 u+u G^{\mu\nu} \notag \\
    &+\frac{1}{2}\big(s^{\alpha\beta}R_{\alpha\beta}g^{\mu\nu}+\nabla_\alpha \nabla^\mu 
    s^{\alpha\nu}+\nabla_\alpha \nabla^\nu s^{\alpha\mu} \notag  \\ & -\nabla^2 
    s^{\mu\nu}-g^{\mu\nu}\nabla_\alpha \nabla_\beta s^{\alpha\beta}\big) +\frac{1}{2}\big(t^{\alpha\beta\gamma\mu}R_{\alpha\beta\gamma}^{\phantom{\alpha\beta\gamma}\nu}  
    \notag \\
    & +t^{\alpha\beta\gamma\nu}R_{\alpha\beta\gamma}^{\phantom{\alpha\beta\gamma}\mu}
    +t^{\alpha\beta\gamma\delta}R_{\alpha\beta\gamma\delta}g^{\mu\nu} \big) \notag
    \\ &-\nabla_\alpha\nabla_\beta t^{\mu\alpha\nu\beta}     -\nabla_\alpha\nabla_\beta t^{\nu\alpha\mu\beta}\,.
\end{align}
According to the approach used to ensure compatibility in the presence of explicit 
diffeomorphism violation, a set of Killing vectors $\xi_i^\mu$ 
associated with the given metric shall be 
identified~\cite{Reyes:2024ywe,Reyes:2024hqi}. It is 
then required that both the energy-momentum tensor of the background 
fields and the background fields themselves remain invariant along these Killing directions, i.e.,
\begin{align}
    \mathcal{L}_{\xi_{i}}T_{\mu\nu}=0 \,,
\end{align}
and 
\begin{align}\label{background-inv}
    \mathcal{L}_{\xi_{i}}u=\mathcal{L}_{\xi_{i}}s^{\mu\nu}
    =\mathcal{L}_{\xi_{i}}t^{\mu\nu\rho\sigma}=0 \,.  
\end{align}
Under these conditions, the Killing vectors associated with the G\"{o}del metric are examined in the following sections. 
%..............................................................................
\section{ The G\"{o}del metric: isometries and form-invariant tensors}\label{Godel}
%..............................................................................
In this section, we introduce the G\"{o}del metric and find 
the two rank tensor that are form-invariant under the 
Killing directions. 
The line element of the G\"{o}del metric is given by
\begin{align}\label{metric}
    \mathrm{d}s^2&=\frac{1}{2\omega^2}\bigg(-\mathrm{d}t^2-2e^x \mathrm{d}t\mathrm{d}y+\mathrm{d}x^2
    \notag \\ &\hspace{2em} -\frac 12  e^{2x}\mathrm{d}y^2+\mathrm{d}z^2\bigg)\,,
\end{align}
where $\omega$ is a constant that represents
the angular velocity of the cosmic rotation \cite{Obukhov}.

The combination of the metric 
and the Einstein field equations with a cosmological 
constant and assuming that the matter content is a pressureless dust, 
the following condition is obtained in order to satisfy the equations
\begin{align}
\Lambda=-\omega^2=-\frac{\kappa\rho}{2} \,. \label{Godel-GR}
\end{align}
In order to investigate the G\"{o}del form-invariant tensors that will serve as background 
fields, it is essential to first examine the symmetries of the G\"{o}del metric 
through its Killing vectors. The G\"{o}del spacetime is a 
highly symmetric solution, admitting five independent Killing vectors. 
These vectors correspond to the spacetime isometries and reflect its homogeneity 
and stationarity. Specifically, three of the Killing vectors are associated with 
translations in time and space, while the remaining two correspond to rotational 
and Lorentz-boost-like symmetries in the spatial sections. These symmetries play 
a crucial role in the physical interpretation of the G\"{o}del 
universe, particularly in relation to its global structure. The 
explicit form of the Killing vectors provide insight into the conserved 
quantities and invariant properties of fields propagating in this background. 
Then, the Killing vectors are
\begin{subequations} \label{Killing-vector-godel}
\begin{align}
    \vec{\xi}_1&= \partial_t\,, \\
    \vec{\xi}_2&=\partial_y \,,  \\
    \vec{\xi}_3&=\partial_z \,, \\
    \vec{\xi}_4&=\partial_x-y\partial_y \,,  \\
    \vec{\xi}_5&=-2e^{-x}\partial_t+y\partial_x+\Big(e^{-2x}-\frac{1}{2}y^2\Big)\partial_y \,.
\end{align}
\end{subequations}
Taking into consideration the conditions in Eq.~\eqref{background-inv}, 
we solve the form-invariant tensor condition for a scalar field $f$, 
a symmetric rank-$2$ tensor $M^{\mu\nu}$, and  a Riemann-like rank-4 tensor $T^{\mu\nu\rho\sigma}$. 

We start considering the first set of Killing vectors $\xi_{i}$, with $i=1,2,3$. 
Since the components of these Killing vectors are constant, the Lie derivative along these directions is given by
\begin{align}
    \mathcal{L}_{\xi_{i}}f&=\xi_{i}^\lambda \partial_\lambda f\,, \\
    \mathcal{L}_{\xi_{i}}M^{\mu\nu}&=\xi_{i}^\lambda \partial_\lambda M^{\mu\nu} \,, \\
    \mathcal{L}_{\xi_{i}}T^{\mu\nu\rho\sigma}&=\xi_{i}^\lambda \partial_\lambda T^{\mu\nu\rho\sigma}\,.
\end{align}
It is then straightforward to verify that each form-invariant 
tensor are independent of the coordinates $t$, $y$, and $z$. 

For the cases $i=4,5$, the situation is different, and the condition 
Eq.~\eqref{background-inv} must be solved explicitly for the specific background under consideration. 

An arbitrary function $f(x)$ gives the equations 
\begin{align}
    \partial_x f(x)=0=y \partial_x f(x) \,,
\end{align}
which leaves only the constant function as a scalar form-invariant for the G\"{o}del metric. 

An arbitrary symmetric rank-2 tensor $M^{\mu\nu}$ is now considered. 
The Lie derivative along the Killing directions is given by:
\begin{eqnarray}
    \mathcal{L}_{\xi_{i}}M^{\mu\nu}=\xi_{i}^\lambda \partial_\lambda 
    M^{\mu\nu}-\partial_\lambda \xi_{i}^\mu M^{\lambda\nu}-\partial_\lambda \xi_{i}^\nu M^{\mu\lambda} \,,
\end{eqnarray}
for $i=4,5$. For $i = 4$, we use the explicit form of the Killing vector~\eqref{Killing-vector-godel}, obtaining: 
\begin{eqnarray}
    \mathcal{L}_{\xi_4}M^{\mu\nu}= \partial_1 M^{\mu\nu}+\delta^\mu_2 M^{2\nu}+\delta^\nu_2 M^{\mu 2}=0 \,. \label{killing-4}
\end{eqnarray}
Taking $\mu,\nu=0$ in Eq.~\eqref{killing-4} we obtain the equation
\begin{eqnarray}
    \partial_1 M^{00}=0
\end{eqnarray}
with solution $M^{00}=:M_1$, where $M_1$ is an arbitrary constant. Taking $\mu=0,\nu=2$ we obtain
\begin{eqnarray}
    \partial_1 M^{02}+M^{02}=0
\end{eqnarray}
we obtain $M^{02}=M_3e^{-x}$. 

Solving this equation constraints for all the components of $M^{\mu\nu}$ it takes the form:
\begin{eqnarray}
    M^{\mu\nu}=\left(\begin{matrix} M_1&M_2 & M_3e^{-x}& M_4 \\
    M_2&M_5 & M_6e^{-x}& M_7 \\
   M_3e^{-x} & M_6e^{-x} & M_8e^{-2x}& M_9 e^{-x} \\
    M_4 & M_7 & M_9e^{-x}& M_{10}\end{matrix}\right) \,,
\end{eqnarray}
where $M_{i}$, with $i=1,\dots,10$, are arbitrary constants.

Now, considering $\xi_{i=5}$, we find:
\begin{align}
\mathcal{L}_{\xi_{5}}M^{\mu\nu}&=y\partial_1 M^{\mu\nu}-2e^{-x}(\delta^\mu_0 M^{1\nu}+\delta^\nu_0s^{1\mu})  \notag \\
 &- (\delta^\mu_1 M^{2\nu}+ \delta^\nu_1 M^{2\mu})+2e^{-2x}(\delta^\mu_2 M^{1\nu}+\delta^\nu_2 M^{1\mu})\notag \\
 &+y(\delta^\mu_2 M^{2\nu}  +\delta^\nu_2 M^{2\mu})=0 \,.
\end{align}
This imposes the following conditions:
\begin{align}
    M_2&=M_6=M_7=M_9=0\,, \notag\\
    M_3&=-2M_5 \,, \\
    M_8&=2M_5\,. \notag
\end{align}
Thus the most general symmetric rank-$2$ form-invariant tensor for the G\"{o}del metric is:
\begin{eqnarray}
    M^{\mu\nu}=\left( \begin{matrix} M_1 & 0 & -2M_5 e^{-x} & M_4 \\ 
    0 & M_5 & 0 & 0 \\
    -2 M_5 e^{-x} & 0 & 2M_5 e^{-2x} & 0 \\
    M_4 & 0 & 0 & M_{10}\end{matrix}\right) \label{M_inv}\,,
\end{eqnarray}
where $M_{j}$ becomes four arbitrary constants. 

Finally, a rank-4 tensor $T^{\mu\nu\rho\sigma}$ possessing the symmetries of a Riemann tensor is considered. Its Lie derivative along the Killing directions is given by:
\begin{align}
    \mathcal{L}_{\xi_i}T^{\mu\nu\rho\sigma}&=\xi_{i}^\lambda \nabla_\lambda T^{\mu\nu\rho\sigma}-\nabla_\lambda \xi_{i}^\mu T^{\lambda\nu\rho\sigma}-\nabla_\lambda \xi_{i}^\nu T^{\mu\lambda\rho\sigma} \notag  \\ &-\nabla_\lambda \xi_{i}^\rho T^{\mu\nu\lambda\sigma}   -\nabla_\lambda \xi_{i}^\sigma T^{\mu\nu\rho\lambda} \,,
\end{align}
for $i=1,\cdots,5$. Knowing that $T^{\mu\nu\rho\sigma}$ can only depend on the $x$ coordinate and proceeding similarly for $i = 4, 5$, we find that the non-vanishing components of the rank-4 form-invariant tensor are:
\begin{eqnarray}
    T^{txtx}&=& T_1, \quad\quad\quad\quad\quad\quad\quad\quad T^{txtz}=T_2,\nonumber \\
    T^{txxy}&=&T_3 e^{-x},\quad\quad\quad\quad\quad\quad\,\, T^{txxz}=T_4,\nonumber \\   
    T^{txyz}&=&T_5 e^{-x}, \quad\quad\quad\quad\quad\quad \,\,\, t^{tyty}=2(T_1-T_3)e^{-2x},\nonumber \\
    T^{tytz}&=&-2T_4 e^{-x}, \quad\quad\quad\quad\quad \, T^{tyxz}=-T_5 e^{-x},\nonumber \\
    T^{tyyz}&=&2T_4e^{-2x},   \quad\quad\quad\quad\quad\,\,\,\, T^{tztz}=T_6,\nonumber \\
    T^{tzxy}&=&(T_5+T_2)e^{-x}, \quad\quad\quad\,\, T^{tzyz}=T_7 e^{-x},\nonumber \\
    T^{xyxy}&=&T_3 e^{-2x},\quad\quad\quad\quad\quad\,\,\,\,\,  T^{xzxz}=-\frac{1}{2}T_7,\nonumber \\
    T^{yzyz}&=&-T_7 e^{-2x} \,,\label{t-invariant}
\end{eqnarray} 
where $T_{i}$, with $i=1,\cdots,7$, are arbitrary constants. 

From~\eqref{M_inv} using the metric and considering
 new constants $\widetilde{M}_i$ functions of $M_i$ and $\omega$, we write 
\begin{eqnarray}
    M_{\mu\nu}=\left( \begin{matrix} \widetilde{M}_1 &0&\tilde{M}_1e^x &\widetilde{M}_2 \\ 0 & \tilde{M}_3&0&0\\ \widetilde{M}_1e^x&0&\frac{1}{2}(2\widetilde{M}_1+\widetilde{M}_3)e^{2x} &\widetilde{M}_2e^x \\ \widetilde{M}_2&0&\widetilde{M}_2e^x & \widetilde{M}_{4}\end{matrix}\right) \,.
\end{eqnarray}

In the next section, these form-invariant tensors will 
serve as the background fields $u$, $s^{\mu\nu}$, and $t^{\mu\nu\rho\sigma}$ 
that characterize Lorentz-violating effects in the gravitational sector of 
the SME. 
%...............................................................................................................
\section{The G\"{o}del universe in the presence of backgrounds}\label{sec:implementation}
%...............................................................................................................
In this section, we investigate the G\"{o}del solution within the gravitational 
sector of the SME in a cosmological context. 
As shown in Eq.~(\ref{EM}), the modified gravitational field equations are given by
\begin{eqnarray}
    G_{\mu\nu}+\Lambda g_{\mu\nu}=(T^{Rstu})_{\mu\nu}+\kappa (T_m)_{\mu\nu}\,.\label{21}
\end{eqnarray}
Here, the dust is considered as the matter content, whose energy-momentum tensor is defined as
\begin{eqnarray}
    (T_m)_{\mu\nu}=\rho u_\mu u_\nu \,,
\end{eqnarray}
where $\rho$ is the energy density and $u_\mu$ is the four-velocity. 
Considering the fluid rest frame, the four-velocity is defined as
\begin{eqnarray}
u^{\mu}=(u^0,0,0,0)\,.
\end{eqnarray}
Using the normalization condition $g_{\mu\nu}u^\mu u^\nu=-1$ and the metric (\ref{metric}), we find
\begin{eqnarray}
u^{\mu}=(\sqrt{2}\,\omega,0,0,0)\,.
\end{eqnarray}
Thus, the covariant components become
\begin{eqnarray}
u_{\mu}=\left(-\frac{1}{\sqrt{2}\,\omega},0,-\frac{e^x}{\sqrt{2}\,\omega},0\right)\,.
\end{eqnarray}
With these elements, the matter source in the matrix form becomes
\begin{eqnarray}\label{dust}
    (T_m)_{\mu\nu}&=&\frac{\rho}{2\omega^2}\left(\begin{matrix}
    1&0&e^x&0\\0&0&0&0\\e^x&0&e^{2x}&0\\0&0&0&0\end{matrix}\right)=\frac{\rho}{2\omega^2}R_{\mu\nu} \,,
\end{eqnarray}
with $R_{\mu\nu}$ being the Ricci tensor for the G\"{o}del metric.
Now, let us analyze Eq.~(\ref{21}) by examining the energy-momentum tensor $(T^{Rstu})_{\mu\nu}$ in terms of its individual sector contributions. Specifically, we consider each sector separately: for the $u$-sector, the tensor becomes $(T^{Ru})_{\mu\nu}$; for the $s$-sector, $(T^{Rs})_{\mu\nu}$; and for the $t$-sector, $(T^{Rt})_{\mu\nu}$.
%...............................................................................................................
\subsection{$u$ sector}\label{u}
%...............................................................................................................
It is found that the energy-momentum tensor associated with the 
$u$-sector, for a G\"{o}del form-invariant 
scalar (which are essentially constant), is given by
\begin{eqnarray}
    (T^{Ru})_{\mu\nu}=\frac{u}{2}\left( \begin{matrix} 1&0&e^x&0 \\ 0&1&0&0
    \\ e^x & 0 &\frac{3}{2}e^{2x}&0 \\ 0&0&0&1\end{matrix} \right)=u G_{\mu\nu}\,.\label{Tu}
\end{eqnarray}
Note that in this simpler case the no-go constraint 
is automatically satisfied since $(T^{Ru})^{\mu\nu}$ is proportional 
to the Einstein tensor and the scalar form-invariant $u$ is a constant. 
The property of satisfying Eq.~\eqref{eq:consistency-requirement} ensures the 
compatibility between the geometry and dynamics and it must be verified in each sector.

Thus, the Einstein field equations become
\begin{eqnarray}
    G_{\mu\nu}+\Lambda g_{\mu\nu}&=& (T^{Ru})_{\mu\nu}+\kappa(T_m)_{\mu\nu}\,.
\end{eqnarray}
Using the metric~(\ref{metric}), the energy-momentum tensor for the dust~(\ref{dust}), 
and the energy-momentum tensor for the $u$-sector~(\ref{Tu}), the field equations are given by
\begin{eqnarray}
    \frac{1}{2}(1-u)-\frac{\Lambda}{2\omega^2}&=&\frac{\kappa\rho}{2\omega^2}\,, \\
    \frac{1}{2}(1-u)+\frac{\Lambda}{2\omega^2}&=&0\,, \\
    \frac{3}{4}(1-u)-\frac{\Lambda}{4\omega^2}&=&\frac{\kappa\rho}{2\omega^2}\,. 
\end{eqnarray}
Solving this system of equations we obtain for the energy density
\begin{eqnarray}
\kappa\rho=2\omega^2(1-u)\,,
\end{eqnarray}
and for the cosmological constant
\begin{eqnarray}
\Lambda=-\omega^2(1-u)=-\frac{\kappa\rho}{2}\,.
\end{eqnarray}
It is important to note that the form-invariant background 
field $u$ acts as a scaling factor for the quantities
\begin{eqnarray}
    \Lambda &\rightarrow& \Lambda'=\frac{\Lambda}{1-u}\,, \\
    \rho&\rightarrow& \rho'=\frac{\rho}{1-u}\,.
\end{eqnarray}
Then, it is possible to write 
\begin{eqnarray}
\Lambda'=-\omega^2=-\frac{\kappa\rho'}{2}\,.
\end{eqnarray}
This equation shares the same structure as its counterpart in general relativity, 
with the G\"{o}del universe remaining a valid solution within this Lorentz-violating sector. 
Consequently, the standard results of general 
relativity for the G\"{o}del solution are recovered in the limit $u \to 0$.
%..................................................................................
\subsection{$s$ sector}\label{s}
%..................................................................................
By requiring that the background field $s^{\mu\nu}$ be invariant under 
transformations along the Killing directions of the G\"{o}del metric, 
it is found that $s^{\mu\nu}$ must take the following form
\begin{eqnarray}
    s^{\mu\nu}&=&\left( \begin{matrix} s_1 & 0 & -2s_2 e^{-x} & s_3 \\ 
    0 & s_2 & 0 & 0 \\
    -2 s_2 e^{-x} & 0 & 2s_2 e^{-2x} & 0 \\
    s_3 & 0 & 0 & s_4\end{matrix}\right),
\end{eqnarray}
where $s_i$ with $i=1,\cdots, 4$ are arbitrary constants.

Using this result, the energy-momentum tensor contribution arising from the $s$-sector, 
associated with the form-invariant background field, is given by
\begin{widetext}
\begin{eqnarray}
    (T^{Rs})_{\mu\nu}=\frac{1}{2\omega^2}\left(\begin{matrix}
    \frac{1}{2}\big(6s_2-5s_1\big)&0&\frac{e^x}{2}\big(6s_2-5s_1\big) & s_3 \\
    0&-\frac{s_1}{2} & 0 & 0 \\
    \frac{e^x}{2}\big(6s_2-5s_1\big)&0&\frac{e^{2x}}{4}\big(12s_2-11s_1\big)&s_3 e^x \\
    s_3&0&s_3 e^x&-\frac{1}{2}\big(2s_2-s_1\big)\end{matrix} \right)\,.
\end{eqnarray}
\end{widetext}
In this case, considering $(T^{Rs})_{\mu\nu}$, the Einstein field equations take the form
\begin{eqnarray}
    G_{\mu\nu}+\Lambda g_{\mu\nu}&=& (T^{Rs})_{\mu\nu}+\kappa(T_m)_{\mu\nu}\,.\label{39}
\end{eqnarray}
In terms of components, the field equations are given by the following set
\begin{align}
    \frac{1}{2}-\frac{\Lambda}{2\omega^2}&= \frac{1}{4\omega^2}(6s_2-5s_1)
    +\frac{\kappa\rho}{2\omega^2}\,,  \label{1-s}\\
    0&=s_3\,, \label{2-s}\\
    \frac{1}{2}+\frac{\Lambda}{2\omega^2}&= -\frac{s_1}{4\omega^2} 
    \,, \label{3-s}  \\
    \frac{3}{4}-\frac{\Lambda}{4\omega^2}&= 
    \frac{1}{8\omega^2}(12s_2-11s_1)+\frac{\kappa\rho}{2\omega^2}\,,\label{4-s} \\
    \frac{1}{2}+\frac{\Lambda}{2\omega^2}&= 
    -\frac{1}{4\omega^2}(2s_2-s_1)\,.  \label{5-s}
\end{align}
From Eq. \eqref{2-s}, we know that $s_3=0$. By 
considering Eqs. \eqref{3-s} and \eqref{5-s}, we conclude that
\begin{eqnarray}
    s_1=s_2=s \,, 
\end{eqnarray}
where $s$ is a constant.
It follows that the equations reduce to
\begin{eqnarray}
    \frac{1}{2}-\frac{\Lambda}{2\omega^2}&=& \frac{s}{4\omega^2} 
    +\frac{\kappa\rho}{2\omega^2}\,, \label{6-s}\\
    \frac{1}{2}+\frac{\Lambda}{2\omega^2}&=& -\frac{s}{4\omega^2}\,, 
    \label{7-s} \\
    \frac{3}{4}-\frac{\Lambda}{4\omega^2}&=&  \frac{s}{8\omega^2}
    +\frac{\kappa\rho}{2\omega^2}\,.\label{8-s}
\end{eqnarray}
From Eq. \eqref{7-s}, the following relation is obtained
\begin{eqnarray}
\omega^2 &=& -\Lambda - \frac{s}{2} \,.
\end{eqnarray}
By using Eq. \eqref{7-s} to eliminate $\Lambda$ from Eqs. \eqref{6-s} 
and \eqref{8-s}, the following expression is find
\begin{eqnarray}
     \omega^2&=& \frac{\kappa\rho}{2} \,.
\end{eqnarray}
We observe that the tensor $s^{\mu\nu}$ contributes to the 
cosmological constant while leaving the energy density unchanged. 
This occurs because, under the conditions $s_3 = 0$ and $s_1 = s_2 = s$, we obtain
\begin{align}
    (T^{Rs})_{\mu\nu}=-\frac{s}{4\omega^2}\left(\begin{matrix}-1&0&-e^x & 0 \\
    0&1 & 0 & 0 \\
    -e^x&0&-\frac{1}{2}e^{2x}&0 \\
    0&0&0&1\end{matrix} \right)=-\frac{s}{2}g_{\mu\nu} \,.
\end{align}
Under this consideration, the system of equations given in Eq. (\ref{39}) can be expressed as
\begin{eqnarray}
    G_{\mu\nu}+\bigg(\Lambda+\frac{s}{2}\bigg)g_{\mu\nu}=\kappa(T_m)_{\mu\nu} \,.
\end{eqnarray}
This leads to the relation
\begin{eqnarray}
    -\bigg(\Lambda+\frac{s}{2}\bigg)=\omega^2=\frac{\kappa\rho}{2}\,,
\end{eqnarray}
which connects the cosmological constant $\Lambda$, the parameters $s$ and 
$\omega^2$, and the energy density $\rho$. This result demonstrates that, 
although the cosmological constant is modified by the Lorentz-violating 
background, the G\"{o}del metric remains a consistent and exact solution 
within the $s$-sector of the gravitational sector of the SME, just as it is in general relativity.

Notice that after the dynamics have been imposed we can see that 
Eq.~\eqref{eq:consistency-requirement} is identically satisfied 
by $(T^{Rs})^{\mu\nu}$. This differs from the $u$ sector analysis 
where the condition was fulfilled before solving the dynamics algebraically.
%.............................................
\subsection{$t$ sector}\label{t}
%.............................................
Here, the proposal is to investigate the consistency of the G\"{o}del 
metric within the $t$-sector. Considering the rank-4 form-invariant tensor
found in~\eqref{t-invariant} as background field, it yields an energy-momentum tensor that takes the form
\begin{widetext}
\begin{eqnarray}
    (T^{Rt})_{\mu\nu}=\frac{1}{2\omega^4}\left(\begin{matrix}
    -5T_1+\frac{13}{4}T_3&0&\big(-5T_1+\frac{13}{4}T_3\big)e^x & -2T_4 \\
    0&-\frac{1}{4}T_3 & 0 & 0 \\
    \big(-5T_1+\frac{13}{4}T_3\big)e^x&0&\big(-5T_1+\frac{25}{8}T_3\big)e^{2x}&-2 T_4e^x \\
     -2T_4&0&-2 T_4e^x&T_1-\frac{3}{4}T_3\end{matrix} \right) \,,
\end{eqnarray}
\end{widetext}
where $T_{i}$, with $i = 1, \dots, 7$, are arbitrary constants. Thus, the Einstein field equations become
\begin{eqnarray}
    G_{\mu\nu}+\Lambda g_{\mu\nu}&=& (T^{Rt})_{\mu\nu}+\kappa(T_m)_{\mu\nu}\,.
\end{eqnarray}
When expressed in terms of its independent components, the following is obtained
\begin{align}
    \frac{1}{2}-\frac{\Lambda}{2\omega^2}&=\frac{1}{2\omega^4}(-5T_1+\frac{13}{4}T_3)+\frac{\kappa\rho}{2\omega^2}\,, \label{56}
 \end{align}   
    \begin{align}
    0&=-{2T_4}, \label{57}\\
    \frac{1}{2}+\frac{\Lambda}{2\omega^2}&=-\frac{1}{2\omega^4}\frac{T_3}{4}\,, \label{58} \\
    \frac{3}{4}-\frac{\Lambda}{4\omega^2}&= \frac{1}{2\omega^4}(-5T_1+\frac{25}{8}T_3)+\frac{\kappa\rho}{2\omega^2}\,, \label{59}\\
    \frac{1}{2}+\frac{\Lambda}{2\omega^2}&=\frac{1}{2\omega^4}(T_1-\frac{3}{4}T_3)\,.\label{60}
\end{align}
Here, we observe from Eq.~\eqref{57} that $T_4 = 0$. By applying Eqs.~\eqref{58} and~\eqref{60}, we obtain the following additional constraints
\begin{align}
   2T_1&=T_3 \,.
\end{align}
Using this result, the remaining equations become
\begin{align}
    \frac{1}{2}-\frac{\Lambda}{2\omega^2}&=\frac{3}{4}\frac{T_1}{\omega^4}+\frac{\kappa\rho}{2\omega^2}\,, \label{62} 
   \end{align} 
 \begin{align}   
    1+\frac{\Lambda}{\omega^2}&=-\frac{T_1}{2\omega^4}\,,  \label{63}
  \end{align}  
   \begin{align}  
    \frac{3}{4}-\frac{\Lambda}{4\omega^2}&=\frac{5}{8} \frac{T_1}{\omega^4}+\frac{\kappa\rho}{2\omega^2}\,.\label{64}
\end{align}
From this set of equations, a system can be formulated 
to express $\omega^2$ in terms of $\Lambda$ or $\rho$, i.e.,
\begin{eqnarray}
 \omega^4+\Lambda \omega^2+\frac{T_1}{2}&=& 0\,, \\
  \omega^4-\frac{\kappa\rho}{2}\omega^2-\frac{T_1}{2} &=&0\,.
\end{eqnarray}
Here, it is recognized that setting $T_1 = 0$ leads to the standard result from general relativity Eq.~\eqref{Godel-GR}. When this expression is solved for $\omega^2$, the following is obtained
\begin{align}
  - \frac{\Lambda}{2}\pm\frac{1}{2}\sqrt{\Lambda^2-2T_1} = \omega^2=\frac{\kappa\rho}{4}\pm \frac{1}{2}\sqrt{\bigg(\frac{\kappa\rho}{2}\bigg)^2+2T_1}\,.
\end{align}
It is important to note that, in order to connect this result with the case $T_1 = 0$ — that is, the standard result from general relativity — we must choose the positive square root in the equation above. Then,
\begin{align}
   -\frac{\Lambda}{2}-\frac{1}{2}\sqrt{\Lambda^2-2T_1} = \omega^2=\frac{\kappa\rho}{4}+ \frac{1}{2}\sqrt{\bigg(\frac{\kappa\rho}{2}\bigg)^2+2T_1}\,. 
\end{align}
Therefore, our result demonstrates that the $t$-sector admits the G\"{o}del metric as an exact solution, although the relation between the cosmological constant and the energy density is modified by the presence of the background field.
After the dynamics have been imposed we can see that the energy-momentum of the background field becomes
\begin{widetext}
\begin{eqnarray}
    (T^{Rt})_{\mu\nu}=\frac{T_1}{4\omega^4}\left(\begin{matrix} 3&0&3e^x & 0 \\
    0&-1 & 0 & 0 \\
    3e^x&0& \frac{5}{2} e^{2x}&0 \\
     0&0&0&-1\end{matrix} \right)=\frac{T_1}{2\omega^4}\big(G_{\mu\nu}-2\omega^2 g_{\mu\nu}\big)\,,
\end{eqnarray}
\end{widetext}
This shows that as in the $s$-sector case, the energy-momentum tensor for the background field $t$ satisfies the consistency requirement Eq.~\eqref{eq:consistency-requirement} only after dynamics is imposed. 

Up to this point, the consistency between the G\"{o}del solution and all sectors describing explicit diffeomorphism violation has been established. 
%..........................................................
\section{On Causality and its violation}\label{causality}
%..........................................................
In the previous section, it was shown that the G\"{o}del metric constitutes a consistent solution within a gravitational theory that includes background fields leading to explicit violations of diffeomorphism invariance and Lorentz symmetry. In this section, attention is turned to another fundamental feature of the G\"{o}del solution: the existence of CTCs, which imply the possibility of causality violation. Within this context, the computation of the critical radius associated with the solution is essential, as causality is violated beyond this radius.

A remarkable feature of this solution is that the existence of CTCs is permitted, implying the theoretical possibility of time travel and raising profound questions about causality~\cite{Godel}.

To determine the critical radius, the metric given in Eq. (\ref{metric}) is expressed in cylindrical coordinates, and a condition is imposed on the $g_{\phi\phi}$ component. A circular curve defined by $C=\{(t,r,\phi, z); t, r, z =\,const; \phi \in [0,2\pi]\}$ represents a CTC if $g_{\phi\phi}$ becomes negative for certain values of $r$ \cite{Godel, Reboucas}. This condition indicates the existence of a noncausal region for $r>r_c$, where the critical radius is given by
\begin{eqnarray}
r_c=\frac{\alpha}{\omega}  \,,
\end{eqnarray}
where $\alpha=\sqrt{2}\,\sinh^{-1}(1)$.

It is important to emphasize that the determination of the critical radius is based solely on the properties of the metric and is therefore independent of the specific gravitational theory under consideration. However, it does depend on the parameter $\omega$, which characterizes the rotation of the G\"{o}del universe. While the general structure of the condition for the critical radius remains the same across different theories, the value of $\omega$ can be modified by the presence of additional terms or background fields specific to each framework. As a result, the critical radius may vary depending on the underlying theory. In this context, we now investigate how the background fields introduced in each sector of the gravitational theory discussed in the previous sections influence the value of $\omega$, and consequently, the critical radius.

Here, each sector will be considered separately in order to determine the corresponding critical radius. 

In the $u$ sector, Sec.~\ref{u}, after the set of field equations was solved, the following result was obtained
\begin{eqnarray}
\omega^2=\frac{\kappa\rho}{2(1-u)}  \,.
\end{eqnarray}
This leads to the critical radius
\begin{eqnarray}
r_c=\alpha\sqrt{\frac{2(1-u)}{\kappa\rho}}\,.
\end{eqnarray}
It is shown that the background field $u$ 
acts to reduce the critical radius. Consequently, 
the critical radius remains finite, implying that causality violation persists.

In the $s$ sector, Sec.~\ref{s}, has been obtained that
\begin{eqnarray}
  \omega^2=  -\bigg(\Lambda+\frac{s}{2}\bigg)\,,
\end{eqnarray}
as a consequence the critical radius becomes
\begin{eqnarray}
r_c=\alpha\sqrt{\frac{2}{-(2\Lambda+s)}}\,.
\end{eqnarray}
In order to obtain a finite and real value, a new condition is imposed on the parameter $s$ and the cosmological constant, namely $2\Lambda+s<0$. If this condition is satisfied, causality violation is permitted in this sector. Another important observation arises at this point: if $s=-2\Lambda$, the critical radius becomes infinite, and causality violation is avoided, resulting in a fully causal region. Therefore, in the $s$ sector, both causal and non-causal regions may emerge, depending on the relationship between the parameter $s$ and the cosmological constant $\Lambda$.

Now, the $t$ sector, studied in 
Sec.~\ref{t}, is analyzed, where the following result has been found
\begin{eqnarray}
\omega^2=\frac{\kappa\rho}{4}+ \frac{1}{2}\sqrt{\bigg(\frac{\kappa\rho}{2}\bigg)^2+2T_1}\,.
\end{eqnarray}
In this case, the critical radius is given by the expression
\begin{eqnarray}
r_c=\alpha\sqrt{\frac{2}{\frac{\kappa\rho}{2}
+\left[\left(\frac{\kappa\rho}{2}\right)^2+2T_1\right]^{1/2}}}\,.
\end{eqnarray}
It is important to note that the presence of the parameter $T_1$ increases the denominator of the expression for the critical radius. As a result, the value of the critical radius is reduced. Therefore, in this sector, a finite critical radius is obtained, indicating that a non-causal region is permitted and causality violation remains possible.
%.................................................
\section{Conclusions}\label{sec:conclusions}
%.................................................
In this work, explicit violations of diffeomorphism 
and Lorentz invariance were considered within 
the framework of the minimal gravitational sector of the SME and the G\"{o}del universe. 
The main objective was to investigate the consistency of the G\"{o}del universe in this modified gravitational theory. Since the G\"{o}del metric is an exact solution in GR, any viable alternative theory of gravity should, ideally, accommodate such solutions under appropriate conditions. To this end, the Killing vectors associated with the G\"{o}del solution were first analyzed, and the G\"{o}del invariants corresponding to all sectors of the modified theory were subsequently computed. These invariants serve as non-dynamical background fields and contribute to the modification of the Einstein field equations.
A universe filled with pressureless dust and a cosmological constant was then 
considered as the matter content. Under these conditions, it was shown that the 
G\"{o}del metric remains a consistent solution in all sectors of the SME gravitational 
theory. However, the specific conditions required to solve the modified field equations 
were shown to depend on the background fields, leading to deviations from the standard GR case.

A crucial part was also to revisit the Noether theorem in the presence of background fields. We considered
an electromagnetic model and CPT-odd term with a CS spacetime-dependent term. We have made a parallel
with the problems one encounters in the case of gravity.  We have identified a sort of 
no-go constraint
that is required for consistency and which may not have solution. 
At the light of the Noether theorem we have 
seen that a constant background produces a conserved current, however in the case the background
is space dependent and imposed to 
be a gradient, the symmetry 
can be restored using a type of St\"{u}ckelberg approach.
We have considered gravity in the presence of a background. In this case we have seen 
that new Noether identities that involve the backgrounds arise. 
We have presented an example in the $u$ sector of the SME and derived the exact 
conserved current associated in the model, identifying a new Komar-like current.

In addition, the question of causality and its potential violation was addressed through the analysis of CTCs. It was demonstrated that modifications introduced by the background fields alter the critical radius that determines the boundary between causal and non-causal regions. The results reveal that, depending on the sector considered, the background fields either increase or decrease this radius, allowing for the existence of both causal and non-causal regions. This represents a novel result: for the same matter content, the presence of background fields in the gravitational SME allows for different causal structures. In particular, sectors were identified in which causality violation can be avoided, while in others, non-causal regions persist. These findings highlight the sensitivity of G\"{o}del solution to Lorentz and diffeomorphism-violating effects and provide further insight into the rich phenomenology of the gravitational SME.
%.......................................
\section{Acknowledgments}
%.......................................
CR acknowledges support from the ANID fellowship No. 21211384 and Universidad de Concepci\'on. 
CMR acknowledges support to the Fondecyt Regular project No. 1241369.
The work of AFS is partially supported by National Council for Scientific and Technological
Development-CNPq project No. 312406/2023-1.
\appendix
%...................................................................
\section{Noether identities in the $u$ sector of the SME}\label{Appendix}
%....................................................................
In this section we derive the Noether identities in the presence of the background $u$
of the minimal gravitational sector of the SME.

We start with the action 
\begin{align}
    S_{\text{EH,u}}=\int\mathrm{d}^4x \;\sqrt{-g}(1-u) R\,,
\end{align}
where $u$ is a background~\cite{Kostelecky:2003fs}.

For a symmetry variation we obtain
\begin{align}\label{var1.}
    \delta S_{\text{EH,u}}[g_{\mu \nu}, \delta_{\xi}
    g_{\mu \nu}]&=-\int\mathrm{d}^4x\, \sqrt{-g}\notag  \\ & 
    \phantom{{}={}}\times \nabla_{\mu} \big( \xi^{\mu}\left(1-u\right) R\big)\,,
\end{align}
and for an arbitrary variation of the fields, including the background
\begin{align}
    \delta S_{\text{EH,u}}[g_{\mu \nu}, \delta g_{\mu \nu}]  &=\int\mathrm{d}^4x 
    \Big[ \sqrt{-g}R \delta \left(1-u\right) \notag \\
    &\phantom{{}={}}+\left(1-u\right)\delta \left(\sqrt{-g}  R\right)\Big]\,.
\end{align}
Now, following the Noether method, we set both variations to be 
the same $\delta g_{\mu \nu}= \delta_{\xi} g_{\mu \nu}$.
Also, we consider $\delta_{\xi} g_{\mu \nu}=-\mathcal L_{\xi} g_{\mu \nu} $ and 
$\delta_{\xi} u= -\mathcal L_{\xi}u$, so
we can write 
\begin{align}\label{var2.}
   \delta S_{\text{EH,u}}[g_{\mu \nu}, \delta_{\xi} g_{\mu \nu}] 
   &=-\int\mathrm{d}^4x   \Big[ \sqrt{-g}  R \mathcal L_{\xi} (1-u) \notag \\ 
   &\phantom{{}={}}+(1-u)\mathcal L_{\xi} (\sqrt{-g} R)\Big]\,.
\end{align}
Subtracting~\eqref{var1.} and~\eqref{var2.}, we have
\begin{align}\label{diffS}
   0&=\int\mathrm{d}^4x  \bigg[  -\sqrt{-g} \nabla_{\mu}
   \big( \xi^{\mu}(1-u) R\big) \notag \\& \phantom{{}={}}+ \sqrt{-g}R \mathcal L_{\xi} (1-u) 
   +(1-u)\mathcal L_{\xi} ( \sqrt{-g}R)\bigg]\,.
   \end{align}
Evaluating the Lie derivatives, we arrive at
\begin{align}\label{diffS2}
   0&=\int\mathrm{d}^4x \sqrt{-g}  \bigg[\nabla_{\mu} \left( \xi^{\mu}(1-u) R\right)
   +R \xi^{\mu}\nabla_{\mu }u  \notag \\& \phantom{{}={}} +2(1-u)G_{\mu \nu} \nabla^{\mu}\xi^{\nu}
    - (1-u)P^{\mu \nu \alpha \beta}\notag \\ &\phantom{{}={}}\times \nabla_{\mu} \nabla_{\nu} ( \nabla_{\alpha}\xi_{\beta}+
    \nabla_{\beta}\xi_{\alpha}) \bigg]    \,,
\end{align}
where $P^{\mu \nu \alpha \beta}$ has been defined in~\eqref{def_P}.
We manipulate the third term in~\eqref{diffS2}
\begin{align}
  2(1-u)G_{\mu \nu} \nabla^{\mu}\xi^{\nu}&=
    2 \nabla_{\mu} \big((1-u)G^{\mu}_{\phantom{\mu} \nu} \xi^{\nu} \big)\notag 
     \\  &\phantom{{}={}}- 2\xi^{\nu}  \nabla^{\mu} \big((1-u)G_{\mu \nu}\big)  \,,
\end{align}
and after some cancellations, we obtain
\begin{align}
   0&=\int\mathrm{d}^4x \sqrt{-g}  \bigg[
   R \xi^{\mu}\nabla_{\mu }u +2 \nabla_{\mu} \big((1-u)R^{\mu}_{\phantom{\mu} \nu} \xi^{\nu} \big) \notag \\& \phantom{{}={}}  - 2 \xi^{\nu}\nabla^{\mu} \big((1-u)G_{\mu \nu}\big) 
    - (1-u)P^{\mu \nu \alpha \beta}\notag  \\   &  \phantom{{}={}}  \times \nabla_{\mu} \nabla_{\nu} ( \nabla_{\alpha}\xi_{\beta}+
    \nabla_{\beta}\xi_{\alpha})\bigg]\,.
\end{align}
We use the contracted Bianchi identity
\begin{align}
   0&=\int\mathrm{d}^4x \sqrt{-g}  \bigg[
   R \xi^{\mu}\nabla_{\mu }u +2 \nabla_{\mu} \big((1-u)R^{\mu}_{\phantom{\mu} \nu} \xi^{\nu} \big) \notag  \\&  \phantom{{}={}} + 2\xi^{\nu} \nabla^{\mu} (uG_{\mu \nu}) 
    - (1-u)P^{\mu \nu \alpha \beta}\notag  \\   &  \phantom{{}={}} \times  \nabla_{\mu} \nabla_{\nu} ( \nabla_{\alpha}\xi_{\beta}+
    \nabla_{\beta}\xi_{\alpha})\bigg]  \,,
\end{align}
arriving at 
\begin{align}
   0&=\int\mathrm{d}^4x \sqrt{-g}  \Big[2 \nabla_{\mu}J_K^{\mu}+
   R \xi^{\mu}\nabla_{\mu }u\notag  \\ &\phantom{{}={}} -2 \nabla_{\mu} \big(uR^{\mu}_{\phantom{\mu} \nu} \xi^{\nu} \big )    + 2 \nabla^{\mu} \big ( u G_{\mu \nu}\big) \xi^{\nu}
   \notag \\& \phantom{{}={}} +uP^{\mu \nu \alpha \beta} \nabla_{\mu} \nabla_{\nu} ( \nabla_{\alpha}\xi_{\beta}+
    \nabla_{\beta}\xi_{\alpha})\Big]\,,
\end{align}
where we have introduced the Komar current defined in~\eqref{Kom_current}.
We use the standard GR
identically Noether identity 
\begin{align}
 \nabla_{\mu}J_K^{\mu}= 0    \,,
\end{align}
to write
\begin{align}
   0&=\int\mathrm{d}^4x \sqrt{-g}  \bigg[
   R \xi^{\mu}\nabla_{\mu }u -2 \nabla_{\mu} (uR^{\mu}_{\phantom{\mu} \nu} \xi^{\nu} )  \\& 
   \phantom{{}={}}+ 2 \xi^{\nu} \nabla^{\mu} (u G_{\mu \nu})
    +uP^{\mu \nu \alpha \beta} \nabla_{\mu} \nabla_{\nu} ( \nabla_{\alpha}\xi_{\beta}+
    \nabla_{\beta}\xi_{\alpha})\bigg]  \notag \,.
\end{align}
For simplicity, we can reinstall the variation $\delta_{\xi} g_{\alpha \beta} $, 
and focus on the last piece 
\begin{align}
     uP^{\mu \nu \alpha \beta} \nabla_{\mu} \nabla_{\nu} \delta_{\xi} g_{\alpha \beta} &=  \nabla_{\mu} 
     (uP^{\mu \nu \alpha \beta} 
      \nabla_{\nu}   \delta_{\xi} g_{\alpha \beta}   )  \notag \\& \phantom{{}={}} - (\nabla_{\mu}u) 
      P^{\mu \nu \alpha \beta}  \nabla_{\nu} \delta_{\xi} g_{\alpha \beta}\,,
  \end{align}
repeating for the internal first term and the last
\begin{align}
     &uP^{\mu \nu \alpha \beta} \nabla_{\mu} \nabla_{\nu} \delta_{\xi} g_{\alpha \beta} 
     =  \nabla_{\mu} \bigg(
    \nabla_{\nu}   (uP^{\mu \nu \alpha \beta} 
       \delta_{\xi} g_{\alpha \beta}   \big)  \notag  \\ &-  (\nabla_{\nu} u)  P^{\mu \nu \alpha \beta} 
       \delta_{\xi} g_{\alpha \beta}  \bigg)   -     \nabla_{\nu}\bigg( (\nabla_{\mu}u) 
       P^{\mu \nu \alpha \beta}  \delta_{\xi} g_{\alpha \beta}\bigg) \notag \\&  +
        ( \nabla_{\nu} \nabla_{\mu}u) P^{\mu \nu \alpha \beta}  \delta_{\xi} g_{\alpha \beta}\,.
  \end{align}
We can arrange the last term above
  \begin{align}
        ( \nabla_{\nu} \nabla_{\mu}u) P^{\mu \nu \alpha \beta}  \delta_{\xi} g_{\alpha \beta}&=-
         ( \nabla_{\nu} \nabla_{\mu}u) P^{\mu \nu \alpha \beta} ( \nabla_{\alpha}\xi_{\beta}+
    \nabla_{\beta}\xi_{\alpha})
    \notag \\ &=-2 ( \nabla_{\nu} \nabla_{\mu}u) P^{\mu \nu \alpha \beta}  \nabla_{\alpha}\xi_{\beta}\notag \\ &=-
    2 ( \nabla_{\alpha} \nabla_{\beta}u) P^{\alpha \beta \mu \nu }  \nabla_{\mu}\xi_{\nu}\,,
      \end{align}
and also
\begin{align}
 2  ( \nabla_{\nu} \nabla_{\mu}u) P^{\mu \nu \alpha \beta}  \nabla_{\mu}\xi_{\nu}&
 = 2 \nabla_{\mu}    \bigg(\big( \nabla_{\alpha} \nabla_{\beta}u) 
 P^{\alpha \beta \mu \nu } \xi_{\nu}    \bigg) \notag \\& 
   -2   \xi_{\nu} P^{\alpha \beta \mu \nu }\nabla_{\mu} \nabla_{\alpha} \nabla_{\beta}u    \,.  
\end{align}
 Finally, we have 
  \begin{align}
    & uP^{\mu \nu \alpha \beta} \nabla_{\mu} \nabla_{\nu} \delta_{\xi} g_{\alpha \beta} 
     =  \nabla_{\mu} \bigg(
    \nabla_{\nu}   (uP^{\mu \nu \alpha \beta} 
       \delta_{\xi} g_{\alpha \beta}   \big) \\ &-  (\nabla_{\nu} u)  P^{\mu \nu \alpha \beta} 
       \delta_{\xi} g_{\alpha \beta}  \bigg)   -     \nabla_{\nu}\bigg( (\nabla_{\mu}u) 
       P^{\mu \nu \alpha \beta}  \delta_{\xi} g_{\alpha \beta}\bigg) \notag  \\& -2 \nabla_{\mu}   
        \bigg(\big( \nabla_{\alpha} \nabla_{\beta}u) P^{\alpha \beta \mu \nu } \xi_{\nu}    \bigg) 
    +2   \xi_{\nu} P^{\alpha \beta \mu \nu }\nabla_{\mu} \nabla_{\alpha} \nabla_{\beta}u  \notag    \,.
  \end{align}
Considering the energy-momentum tensor for the background field $u$
\begin{eqnarray}
    (T^{Ru})^{\mu\nu}=uG^{\mu\nu}-\nabla^\mu\nabla^\nu u +g^{\mu\nu}\Box u\,,
\end{eqnarray}
 derived in~\cite{Kostelecky:2003fs,Bailey:2006fd}, we write
 \begin{eqnarray}
    (T^{Ru})^{\mu\nu}=uG^{\mu\nu}-  P^{\mu \nu \alpha \beta}  \nabla_\alpha\nabla_\beta u \,,
\end{eqnarray}
 and we have
  \begin{align}
   0&=\int\mathrm{d}^4x \sqrt{-g}  \Bigg[
   R \xi^{\mu}\nabla_{\mu }u -2 \nabla_{\mu} (uR^{\mu}_{\phantom{\mu} \nu} \xi^{\nu} ) \notag \\& 
   + 2 \xi_{\nu} \nabla_{\mu}  (T^{Ru})^{\mu\nu}
   -  \nabla_{\mu} \bigg(
    \nabla_{\nu}   (uP^{\mu \nu \alpha \beta} 
       \delta_{\xi} g_{\alpha \beta}   \big)  \notag  \\& -  (\nabla_{\nu} u)  P^{\mu \nu \alpha \beta} 
       \delta_{\xi} g_{\alpha \beta}  \bigg)   +     \nabla_{\nu}\bigg( (\nabla_{\mu}u) 
       P^{\mu \nu \alpha \beta}  \delta_{\xi} g_{\alpha \beta}\bigg) \notag  \\ & +2 \nabla_{\mu}    \bigg(\big( \nabla_{\alpha} \nabla_{\beta}u) 
       P^{\alpha \beta \mu \nu } \xi_{\nu}    \bigg)  \Bigg]   \,.
\end{align}
  We write
  \begin{align}
   0&=\int\mathrm{d}^4x \sqrt{-g}  \Bigg[
   R \xi^{\mu}\nabla_{\mu }u  
   + 2 \xi_{\nu} \nabla_{\mu}  (T^{Ru})^{\mu\nu}
   \notag \\ & \phantom{{}={}}+  \nabla_{\mu}W^{\mu}  \Bigg]   \,,
\end{align}
  where we have defined 
  \begin{align}
  W^{\mu} &= -2  uR^{\mu}_{\phantom{\mu} \nu} \xi^{\nu} -
    \nabla_{\nu}   (uP^{\mu \nu \alpha \beta} 
       \delta_{\xi} g_{\alpha \beta}   \big)  \\&  \phantom{{}={}}+ 2 (\nabla_{\nu} u)  P^{\mu \nu \alpha \beta} 
       \delta_{\xi} g_{\alpha \beta}  +2   
         \big( \nabla_{\alpha} \nabla_{\beta}u) P^{\alpha \beta \mu \nu } \xi_{\nu}  \notag  \,. 
\end{align}
  Let us analyze the divergence term and show that it vanishes. Consider
  \begin{align}
  \nabla_{\mu}W^{\mu} &=  \nabla_{\mu}\bigg[ -2  uR^{\mu}_{\phantom{\mu} \nu} \xi^{\nu} -
    \nabla_{\nu}   (uP^{\mu \nu \alpha \beta} 
       \delta g_{\alpha \beta}   \big)  \\& \phantom{{}={}} + 2 (\nabla_{\nu} u)  P^{\mu \nu \alpha \beta} 
       \delta g_{\alpha \beta}  +2   
         \big( \nabla_{\alpha} \nabla_{\beta}u) P^{\alpha \beta \mu \nu } \xi_{\nu}     \Bigg] \notag \,,
  \end{align}       
which is          
\begin{align}
         \nabla_{\mu}W^{\mu}&= \nabla_{\mu}\bigg[ 
   -2  uR^{\mu}_{\phantom{\mu} \nu} \xi^{\nu} -P^{\alpha \beta \mu \nu } \nabla_{\nu}   (u
       \delta g_{\alpha \beta}   \big)  \\& \phantom{{}={}} + 2P^{\alpha \beta \mu \nu } (\nabla_{\nu} u)  
       \delta g_{\alpha \beta}  +2   P^{\alpha \beta \mu \nu }
         \big( \nabla_{\alpha} \nabla_{\beta}u)  \xi_{\nu}     \Bigg]\notag  \,,
\end{align}    
or       
\begin{align} 
         \nabla_{\mu}W^{\mu}&= \nabla_{\mu}\bigg[ -2  uR^{\mu}_{\phantom{\mu} \nu} \xi^{\nu} 
    - P^{\alpha \beta \mu \nu } (\nabla_{\nu}   u)
       (\nabla_{\alpha} \xi_{\beta}+ \nabla_{\beta} \xi_{\alpha} ) \notag \\ &\phantom{{}={}}+u P^{\alpha \beta \mu \nu } \nabla_{\nu}   (
       (\nabla_{\alpha} \xi_{\beta}+ \nabla_{\beta} \xi_{\alpha} )   \big)\notag  \\ &\phantom{{}={}} +2   P^{\alpha \beta \mu \nu }
         \big( \nabla_{\alpha} \nabla_{\beta}u)  \xi_{\nu}     \Bigg]   \,.
\end{align}
The first and the fourth term combine to form the Komar current
  \begin{align}
\nabla_{\mu}W^{\mu} &= \nabla_{\mu}\bigg[ -2  uJ_K^{\mu}  
    - P^{\alpha \beta \mu \nu }(\nabla_{\nu}   u)
       (\nabla_{\alpha} \xi_{\beta}+ \nabla_{\beta} \xi_{\alpha} ) \notag \\&\phantom{{}={}} +2  P^{\alpha \beta \mu \nu } 
         \big( \nabla_{\alpha} \nabla_{\beta}u)  \xi_{\nu}     \Bigg]   \,.
\end{align}
Evaluating the derivative yields
\begin{align}
&\nabla_{\mu}W^{\mu} =  -2  (\nabla_{\mu}u)J_K^{\mu}  
    - P^{\alpha \beta \mu \nu }(\nabla_{\mu}\nabla_{\nu}   u)
       (\nabla_{\alpha} \xi_{\beta}+ \nabla_{\beta} \xi_{\alpha} ) \notag \\ &- P^{\alpha \beta \mu \nu }(\nabla_{\nu}   u)
      \nabla_{\mu} (\nabla_{\alpha} \xi_{\beta}+ \nabla_{\beta} \xi_{\alpha} )  \\&+2  P^{\alpha \beta \mu \nu } 
         \big( \nabla_{\mu} \nabla_{\alpha} \nabla_{\beta}u)  \xi_{\nu}    +2  P^{\alpha \beta \mu \nu } 
         \big(  \nabla_{\alpha} \nabla_{\beta}u) \nabla_{\mu} \xi_{\nu}   \notag    \,.
\end{align}
The second and the last term cancel, we have
  \begin{align}
\nabla_{\mu}W^{\mu} &=  -2  (\nabla_{\mu}u)J_K^{\mu}  
    - P^{\alpha \beta \mu \nu }(\nabla_{\nu}   u)
     \notag  \\ &\phantom{{}={}} \times \nabla_{\mu} (\nabla_{\alpha} \xi_{\beta}+ 
     \nabla_{\beta} \xi_{\alpha} ) \notag \\&\phantom{{}={}} +2  P^{\alpha \beta \mu \nu } 
         \big( \nabla_{\mu} \nabla_{\alpha} \nabla_{\beta}u)  \xi_{\nu}        \,.
\end{align}
We replace the Komar current and arrive at
\begin{align}
\nabla_{\mu}W^{\mu} &=  -2 R^{\mu} _{\nu} \xi^{\nu} (\nabla_{\mu}u)
  \notag \\& \phantom{{}={}}+2  P^{\alpha \beta \mu \nu } 
         \big( \nabla_{\mu} \nabla_{\alpha} \nabla_{\beta}u)  \xi_{\nu}       \,,
\end{align}
and use 
\begin{align}
  P^{\alpha \beta \mu \nu } 
          \nabla_{\mu} \big(\nabla_{\alpha} \nabla_{\beta}u)  \xi_{\nu}    &=   \nabla_{\mu}( \nabla^{\mu} \nabla^{\nu}u -g^{\mu \nu} \Box u) \xi_{\nu} 
          \notag \\&=   \Box \nabla^{\nu}u - \nabla^{\nu}\Box u) \xi_{\nu}\notag  \\ &= \xi_{\nu}[ \Box, \nabla^{\nu}]u 
         \notag  \\&= \xi_{\nu}R^{\nu}_{\mu} \nabla^{\mu} u\,.
\end{align}
Both terms cancel and obtain finally
  \begin{align}
\nabla_{\mu}W^{\mu} &\equiv 0   \,.
\end{align}

%%%%%%%%%%%%%%%%%%%%%%%%%%%%%%%%%%%%%%%%%%%%%%%%%%%%%%%%%%%%%%%%%%

%\global\long\def\link#1#2{\href{http://eudml.org/#1}{#2}}
% \global\long\def\doi#1#2{\href{http://dx.doi.org/#1}{#2}}
% \global\long\def\arXiv#1#2{\href{http://arxiv.org/abs/#1}{arXiv:#1 [#2]}}
% \global\long\def\arXivOld#1{\href{http://arxiv.org/abs/#1}{arXiv:#1}}

%%%%%%%%%%%%%%%%%%%%%%%%%%%%%%%%%%%%%%%%%%%%%%%%%%%%%%%%%%%%%%%%%%
%.......................................................................

\end{document}